\documentclass[twocolumn]{aastex62}

\usepackage[graphicx]{realboxes} 
\accepted{February 17, 2020}

\shorttitle{Ionization Corrections for the Neutral ISM}
\shortauthors{Hernandez et al.}

\begin{document}

\title{\uppercase{Newly Improved Ionization Corrections for the Neutral Interstellar Medium: Enabling Accurate Abundance Determinations in Star-forming Galaxies throughout the Universe}\footnote{Based on observations made with the Hubble Space Telescope under program ID 15193.}}



\correspondingauthor{Svea Hernandez}
\email{sveash@stsci.edu}

\author{Svea Hernandez}
\affiliation{Space Telescope Science Institute,
3700 San Martin Drive, 
Baltimore, MD 21218, USA}

\author{Alessandra Aloisi}
\affiliation{Space Telescope Science Institute,
3700 San Martin Drive, 
Baltimore, MD 21218, USA}

\author{Bethan L. James}
\affiliation{Space Telescope Science Institute,
3700 San Martin Drive, 
Baltimore, MD 21218, USA}

\author{Gary J. Ferland}
\affiliation{Department of Physics and Astronomy, 
University of Kentucky, 
Lexington, KY 40506, USA}

\author{Andrew J. Fox}
\affiliation{Space Telescope Science Institute,
3700 San Martin Drive, 
Baltimore, MD 21218, USA}


\author{Monica Tosi}
\affiliation{INAF-- OAS Bologna, 
Via Gobetti 93/3, 
I-40129 Bologna, Italy}

\author{Jason Tumlinson}
\affiliation{Space Telescope Science Institute,
3700 San Martin Drive, 
Baltimore, MD 21218, USA}
\affiliation{Center for Astrophysical Sciences, 
Department of Physics and Astronomy, 
The Johns Hopkins University, 
Baltimore, MD 21218, USA}

\begin{abstract}
Studies measuring the chemical abundances of the neutral gas in star-forming galaxies (SFGs) require ionization correction factors (ICFs) to accurately measure their metal contents. In the work presented here we calculate newly improved ICFs for a sample of SFGs. These new corrections include both the contaminating ionized gas along the line of sight (ICF$_{\rm ionized}$) and unaccounted higher ionization stages in the neutral gas (ICF$_{\rm neutral}$). We make use of recently acquired spectroscopic observations taken with the Cosmic Origins Spectrograph (COS) on board Hubble to measure column densities for \ion{Fe}{2} and \ion{Fe}{3}. Using the \ion{Fe}{3}/\ion{Fe}{2} ratios as well as other physical properties (i.e. $\log$[L$_{\rm UV}$], $N$(\ion{H}{1}), T, and $Z$) we generate ad-hoc photoionization models with \texttt{CLOUDY} to quantify the corrections required for each of the targets. We identify a luminosity threshold of $\log$[L$_{\rm UV}$]$\sim$ 40.75 erg s$^{-1}$ above which the ICF$_{\rm neutral}$ values for nitrogen are relatively higher (ICF$_{\rm neutral}=0.05$-0.7) than those for the rest of the elements (ICF$_{\rm neutral}\sim 0.01$). This behaviour indicates that for the high UV luminosity objects, \ion{N}{2} is found in non-negligible quantities in the neutral gas, making these ICF$_{\rm neutral}$ corrections critical for determining the true abundances in the interstellar medium. In addition, we calculate ICFs from a uniform grid of models covering a wide range of physical properties typically observed in studies of SFGs and extragalactic \ion{H}{2} regions. We provide the community with tabulated ICF values for the neutral gas abundances measured from a variety of environments and applicable to chemical studies of the high redshift universe.   
 \end{abstract}



\section{Introduction}\label{sec:intro}
Metallicity measurements of extragalactic environments have proven to be critical for constraining the star formation and evolution of galaxies. \citet{leq79} found that there is a correlation between the mass of a galaxy and its metallicity. This mass-metallicity relation (MZR) has been used extensively to understand galactic winds, star formation episodes and chemical enrichment in star-forming galaxies (SFGs). Abundances in the interstellar medium (ISM) are typically obtained through the analysis of optical and near-infrared (NIR) emission lines observed in \ion{H}{2} regions. \ion{H}{2} regions, however, are associated with relatively recent star formation, therefore the abundances measured from these locations could be enhanced compared to the surrounding ISM \citep{kun86}. Furthermore, these SFGs are known to host large reservoirs of \ion{H}{1}, which can account for as much as 90-95\% of their total baryonic matter \citep{kni00}. Such reservoirs of neutral gas can in principle hide a large portion of the metals. Assessing the relative bias accompanying the abundance measurements from \ion{H}{2} regions is of general importance for obtaining a clear picture of the metal content in SFGs. This can be accomplished by directly studying the chemical composition of the neutral ISM. \par
The metals of the neutral gas in a given galaxy can be studied through the analysis of the absorption in their far ultraviolet (FUV) spectroscopic observations. A common technique is to use bright UV targets within these galaxies as background sources. In such observations, the metals along the line of sight imprint absorption features on the UV continuum of such targets. In principle, this type of analysis can be done using background quasars \citep{wol05}. Such an approach has been applied extensively to study the halos of galaxies at redshifts $z>$ 0.1 \citep{tum11, wer12, wer13}.  However, quasars that are sufficiently bright in the FUV for this type of analysis are rare in the vicinity of galaxies at $z\sim$ 0. A clear advantage of using UV sources within the galaxies over quasars is that the former probes sightlines much closer to the \ion{H}{2} regions, near the core of the galaxy, whereas the latter might bias the abundance studies if any metallicity gradients are present \citep{kob98, ann15}. \par
In ISM abundance studies of SFGs it is critical to take into account both the dust content and ionization effects due to contaminating ionized gas or contributing higher ionization ions  present but not measured in the neutral gas. Dust opacity in SFGs has been investigated amply especially after the discovery of far infrared (FIR)-bright galaxies at intermediate and high redshift \citep{hug98, lil99, bar99} bringing new urgency to understanding dust opacity and emission in SFGs at all redshifts \citep{cal00, lei02, rom14}. In the present work we investigate both types of ionization corrections mentioned above.  \par
In general, when deriving abundance ratios from column density values it is assumed that the dominant ionization state in \ion{H}{1} regions is representative of the total abundance in the neutral gas. It is well known that elements with their first or second ionization potential $>$ 13.6 eV are found as neutral atoms or single-charged ions in the neutral gas cloud, respectively. The former is the case of  O and N; the latter applies to C, Si, S, and Fe. This means that we expect \ion{O}{1}, \ion{N}{1},  \ion{C}{2}, \ion{Si}{2}, \ion{S}{2}, and \ion{Fe}{2} to be the dominant states in the \ion{H}{1} cloud and representative of the abundances of these elements. Given that these ions/atoms can also exist in the ionized cloud and cause an overestimate of the neutral abundance measurements, one must include ionization corrections for such cases. Similarly, higher ionization species (i.e. \ion{Fe}{3}) can in principle exist in the neutral gas cloud. In order to extract accurate chemical abundances one needs to apply both of these corrections to the inferred abundances. \par
In the work described here we look at the amount of ionized gas contaminating the neutral abundance measurements, as well as the amount of higher ionization ions compared to the dominant ion of a certain species dominating in the H I region and estimate ionization correction factors (ICF$_{\rm ionized}$ and ICF$_{\rm neutral}$, respectively) for a variety of galactic environments. This work combines the analysis of Hubble Cosmic Origin Spectrograph (COS) observations with modeling of the ionized and neutral gas using the photoionization code \texttt{CLOUDY} \citep{fer17}. A similar effort of estimating ICFs for the same sample of star-forming galaxies studied here was performed by our group in  \citet{jam14}. However, we note that given the limited wavelength coverage of our previous observations, we were not able to obtain a ratio of \ion{Fe}{3}/\ion{Fe}{2} for each of the targets (only for the SBS 0335-052 galaxy). This \ion{Fe}{3}/\ion{Fe}{2} ratio is a critical indicator of the gas volume density, an essential component for tailoring photoionization models to the physical properties of each of the galaxies. Due to the absence of the \ion{Fe}{3}/\ion{Fe}{2} ratio, in \citet{jam14} we instead take a conservative approach and calculate photoionization models based on the average properties of the full galaxy sample. Here we present the analysis of new HST/COS observations used to measure the \ion{Fe}{3}/\ion{Fe}{2} ratio in each of the galaxies and generate ad-hoc ionization models accounting for the individual physical properties of each object to estimate more accurate ICF values. \par
In Section \ref{sec:obs} we describe the observations and data reduction. Section \ref{sec:ana} provides information about the analysis including the photoionization models. And finally in Section \ref{sec:discussion} and Section \ref{sec:con} we present a brief discussion and our final remarks, respectively.


\section{Observations and Data Reduction}\label{sec:obs}
\textit{HST/COS} observations of eleven bright UV sources (young stellar clusters) in a sample of nine SFGs were carried out over 33 $HST$ orbits, executing as part of the proposal ID 15193 (PI: Aloisi) between April of 2018 and May of 2019. The targets were acquired using a standard peak-up target acquisition in NUV imaging mode. These data use the G130M and G160M gratings centered on $\lambda$= 1222 \r{A} and $\lambda$= 1623 \r{A}, respectively. Such configurations provide a wavelength coverage ranging between $1068-1800$ \r{A} (we point out the wavelength gap between the two gratings of $\sim$60 \r{A}). The COS observations were taken at Lifetime Position 4 (LP4) which provides spectral resolutions on average of $R\sim$15,000 for both gratings \citep{fox18}. \par
\begin{figure*}
   	  \centerline{\includegraphics[scale=0.4]{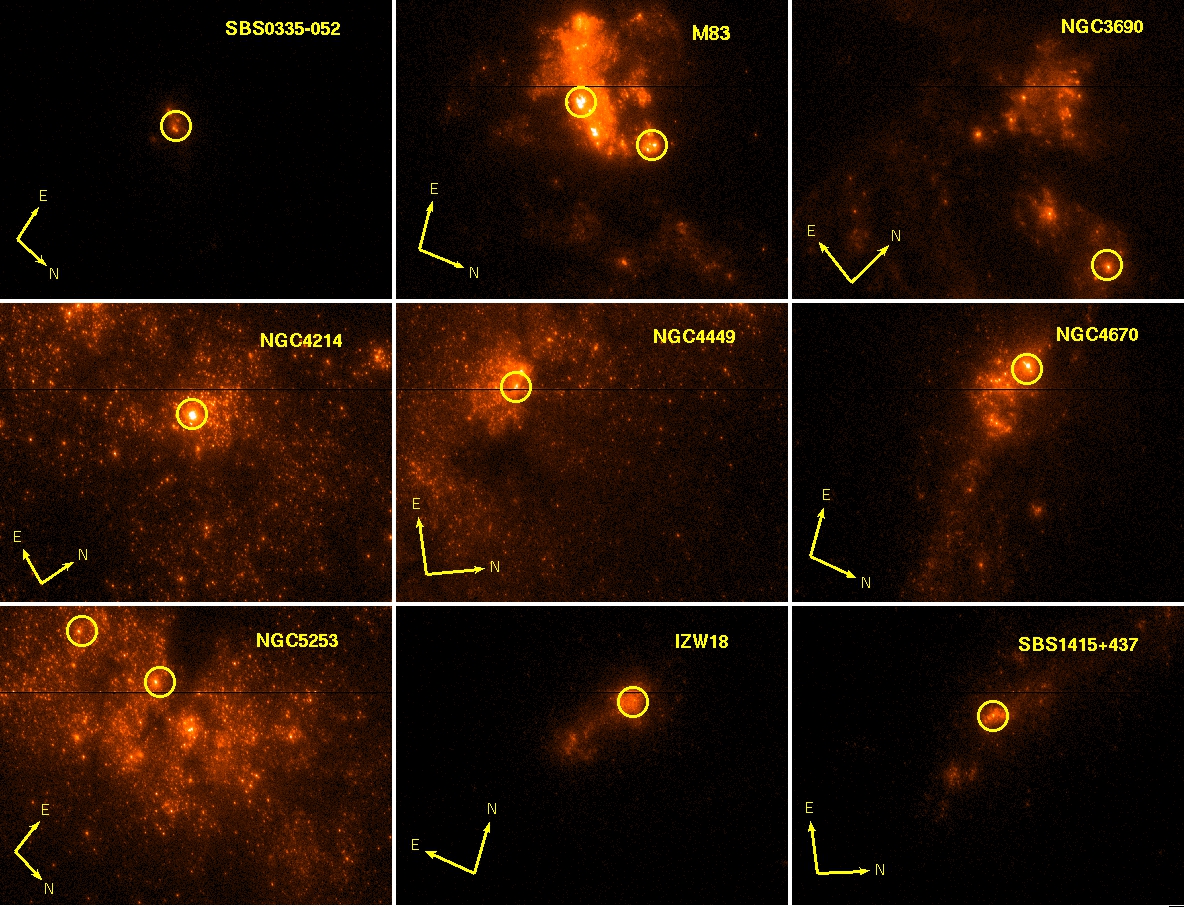}}
      \caption{34\arcsec $\times$ 25\arcsec $ACS/SBC$ mosaic observed with the F125LP filter. The yellow circles display the COS 2.5\arcsec aperture overlaid on the UV bright sources studied here. All of the images were observed as part of PID: 11579 (PI: Aloisi), with the exception of SBS0335-052 which was observed under PID: 9470 (PI: Kunth).}
         \label{fig:sfgs}
   \end{figure*}

Our original galaxy sample of nine SFGs presented in \citet{jam14} was chosen based on the existence of archival $FUSE$ data. Those data were used to confirm our targets were strong emitters in the FUV, and to directly estimate the flux in a similar FUV spectral region as that of $COS$. This sample spans a broad range of metallicities, SFRs, and galaxy spectral types. In Table \ref{tab:gal_prop} we present the basic properties of our galaxy sample ordered by increasing metallicity: the most commonly used name (column 1), the coordinates of the target (columns 2 and 3), galaxy type, metallicity of the galaxy as measured from the O in \ion{H}{2} regions (column 4), radial velocity, distance to the galaxy (column 5), and a list of references used to populate the metallicity column (column 6).  
In Figure \ref{fig:sfgs} we show the ACS/solar blind channel (SBC) FUV imaging of each galaxy in the F125LP filter used to select the COS targets first observed in PID:11579 (PI: Aloisi).\par 
 The new COS observations listed in Table \ref{tab:obs_log} were retrieved from the Mikulski Archive at the Space Telescope Science Institute (MAST). The data were reduced using the standard CALCOS pipeline v.3.3.4.

\begin{table*}
\caption{Basic properties of the SFGs analyzed in this work}
\label{tab:gal_prop}
\centering 
\resizebox{\textwidth}{!}{
\begin{tabular}{cccccccccc}
\hline \hline
Galaxy & R.A. & Decl. & Type$^{a}$ & $E(B-V)^{b}_{\rm MW}$ & $E(B-V)$& 12+$\log$(O/H) & Velocity$^{c}$ & Distance$^{c}$ & Reference$^{d}$  \\
 & (J2000) & (J2000) & & &&   & (km s$^{-1})$ & (Mpc) & 12+$\log$(O/H) \\
\hline\\
I Zw 18 & 09 34 02.298 & $+$55 14 25.07 & BCD & 0.00& 0.032&  7.2 & 753 & 18.2 $\pm$ 1.5 &  1  \\
SBS 0335$-$052 & 03 37 44.002 & $-$05 02 38.32 & BCD &0.04& 0.047& 7.3 & 4,043 & 53.7 $\pm$ 3.8 & 1   \\
SBS 1415$+$437 & 14 17 1.406& $+$43 30 4.75 & BCD &0.01& 0.009& 7.6 & 609 & 13.6 $\pm$ 0.9 & 2  \\
NGC 4214 & 12 15 39.413 & $+$36 19 35.17 & Irr &0.00& 0.022& 8.2 & 291 & 3.04 $\pm$ 0.04 & 3 \\
NGC 5253 & 13 39 56.976 & $-$31 38 27.01 & Irr &0.04& 0.056& 8.2 & 403 & 3.77 $\pm$ 0.20 & 4  \\
NGC 4670 & 12 45 16.906 & $+$27 07 30.02 & BCD &0.01& 0.015& 8.2 & 1,069 & 23.1 $\pm$ 1.6 &  5  \\
NGC 4449 & 12 28 10.816 & $+$44 05 42.95 & Irr & 0.00 & 0.019&8.4 & 203 & 3.82 $\pm$ 0.27 & 6 \\
NGC 3690 & 11 28 31.003 & $+$58 33 41.08 & Merger &0.00& 0.017& 8.8 & 3,119 & 48.5 $\pm$ 3.4 & 5   \\
M83 & 13370.515 & $-$29 52 00.48 & SAB(s)c &0.03 & 0.066& 9.2 & 513 & 4.8 $\pm$ 0.2 & 7\\
\hline
\end{tabular}}
\begin{minipage}{15cm}~\\
\textsuperscript{$a$}{ BCD, blue compact dwarf galaxy; Irr, irregular galaxy; and SAB(s)c, weakly barred spiral galaxy with loosely wound arms (type c) and no ring-like structure.}\\
\textsuperscript{$b$}{MW reddening values from \citet{hec98}.}\\
\textsuperscript{$c$}{Parameters taken from \citet{jam14}.}\\
 \textsuperscript{$d$}{ References: (1) \citet{izo99}, (2) \citet{thu99}, (3) \citet{kob96}, (4) \citet{wal89}, (5) \citet{hec98}, (6) \citet{ann17}, (7) \citet{mar10}.}\\
 \end{minipage}
\end{table*}

\begin{table*}
\caption{$HST/COS$ Observations for program ID: 15193}
\label{tab:obs_log}
\centering 
\resizebox{\textwidth}{!}{
\begin{tabular}{ccccccccc}
\hline \hline
Target & Dataset & R.A. & Decl. & Observation Date & Duration & Setting & S/N$^{a}$ & FWHM$_{\rm spec}^{b}$ \\
 & & (J2000) & (J2000) & (GMT) & (s) &  & & (km s$^{-1}$)  \\
\hline\\
     I Zw 18&	LDN707010&	09 34 01.970&	+55 14 28.10&		2018-11-12 00:49:18&		2352&		G130M/1222&8 & 29\\ 	 
     	 
	SBS 0335-052&	LDN709010&	03 37 43.980&	-05 02 38.90&		2018-08-29 15:46:18	&	4696&		G130M/1222	&10 & 28 \\ 	 	 
	SBS 0335-052&	LDN709020&	03 37 43.980&	-05 02 38.90&		2018-08-29 18:52:07&		5604	&	G160M/1623	&7 & 27 \\ 	 
	
	SBS 1415+437&	LDN708020&	14 17 01.420&	+43 30 05.16&		2018-05-08 09:14:47&		2693	&	G130M/1222	&7 &  34\\  	  	 
	SBS 1415+437&	LDN708010&	14 17 01.420&	+43 30 05.16&		2018-05-08 04:11:27&		2444	&	G160M/1623	&7& 33\\	 
		 	 	  	 	 
	NGC 4214&	LDN703010&	12 15 39.480&	+36 19 35.36&		2018-12-17 13:49:27&		540	&	G130M/1222	&9 &  29 \\ 	 	 
	NGC 4214&	LDN703020&	12 15 39.480&	+36 19 35.36&		2018-12-17 14:07:44&		1156	&	G160M/1623	&13 & 28\\	
	
	NGC 5253-1 & LDN706030 & 13 39 56.020&  -31 38 31.30 & 2019-05-03 12:41:47& 1656 & G130M/1222 & 7 & 42\\
	NGC 5253-1 & LDN706040 & 13 39 56.020& -31 38 31.30	 & 2019-05-03 14:07:46&	3473 & G160M/1623 & 10 & 41\\
	
	NGC 5253-2 & LDN706010 & 13 39 55.889& -31 38 38.34   & 2019-05-03 08:56:30 & 3013 & G130M/1222 & 8 & 30\\
	NGC 5253-2 & LDN706020 & 13 39 55.889& -31 38 38.34  & 2019-05-03 10:43:11 &	3409 & G160M/1623 & 7 & 28\\
	
	NGC 4670&	LDN705010&	12 45 17.265&	+27 07 32.13&		2018-08-07 02:12:33&		400	&	G130M/1222	&5 & 29 \\	 	 	 
	NGC4670&	LDN705020&	12 45 17.265&	+27 07 32.13&		2018-08-07 02:28:30&		1172	&	G160M/1623	&8 &  28\\	 
	
	NGC 4449&LDN754010	&	12 28 11.089&	+44 05 37.06	&	2019-04-04 00:55:09	&	680	&	G130M/1222	&6 &  44\\	 	 	 
	NGC 4449&LDN754020	&	12 28 11.089&	+44 05 37.06	&	2019-04-04 02:23:35	&	964	&	G160M/1623	& 5&  41\\	 
			 	  
	NGC 3690&	LDN752010&	11 28 29.149&	+58 33 41.01&		2018-09-24 15:20:53&		5316	&	G130M/1222	&11 &  29 \\ 	 	 
	NGC 3690&	LDN752020&	11 28 29.149&	+58 33 41.01&		2018-09-24 18:35:31&		6092	&	G160M/1623	&9 & 28 \\	 
	
	M83-1&	LDN701020&	13 37 00.458&	-29 51 54.58&		2018-04-20 18:19:02&		916&		G130M/1222	&8 & 28\\	 		 	 	 
	M83-1&	LDN701010&	13 37 00.458&	-29 51 54.58&		2018-04-20 16:49:39&		1240	&	G160M/1623	 &7 & 27 \\	 	 
	 	 	 	 
	M83-2&	LDN701030&	13 37 00.508&	-29 52 01.22&		2018-04-20 18:50:30&		380&		G130M/1222	&8 & 31\\	 	 	 
	M83-2&	LDN701040&	13 37 00.508&	-29 52 01.22&		2018-04-20 20:04:45&		420&		G160M/1623	 &7 & 30\\	 	 
\hline
\end{tabular}}
\begin{minipage}{15cm}~\\
\textsuperscript{$a$}{ S/N values are estimated at 1350 \r{A} for G130M and at 1500 \r{A} for G160M.}\\
\textsuperscript{$b$}{ FWHM$_{\rm spec}$ estimated at 1150 \r{A} for G130M and at 1450 \r{A} for G160M.}\\
 \end{minipage}
\end{table*}

\section{Analysis}\label{sec:ana}
\subsection{Continuum Fitting}
Similar to the work done in \citet{jam14}, the abundance analysis here relied on the normalized spectra of the different targets. We fit the continuum by interpolating between points free of absorption, which we refer to as \textit{nodes}. The placement of the nodes was carefully chosen after visually inspecting the individual spectra. Our software uses a spline function for interpolating between the manually defined nodes. 

\subsection{Line-profile fitting}
We derive column densities for \ion{Fe}{2} and \ion{Fe}{3} by fitting Voigt profiles to their observed absorption features. We make use of the recently developed Python code \texttt{VoigtFit} v.0.10.3.3 \citep{kro18}. This software was created to handle the fitting of absorption profile features accounting for the line broadening introduced by the spectrograph used to record the observations (i.e. line spread function, LSF). \texttt{VoigtFit} allowed us to convolved the COS LP4 LSF with the intrinsic model profiles (see Section \ref{sec:lsf} for more details on the spectral resolution of the observations). The Python module was optimized for multicomponent fitting allowing for deblending of different components along the line of sight. We validated the new software by re-estimating the column densities published in \citet{jam14} using the old observations and found excellent agreement between the software used in that publication (\texttt{FITLYMAN}, \citealt{fon95}) and this new code.\par
We note that from the COS observations we detect an average absorption arising from the neutral gas along the line of sight of the observed stellar clusters. This implies that the total absorption observed in the FUV spectra comes from a combination of the many unresolved velocity components along the sightlines enclosed by the 2.5\arcsec aperture. Possibly counterintuitively, \citet{jen86} showed that several populations of ISM lines (multiple lines of sight) can indeed be analyzed collectively applying a single-velocity component approximation. \citet{jen86} finds that one can obtain relatively accurate total column densities as long as the equivalent widths of the multiple lines are combined and there is no irregular (i.e. bimodal) distribution function for the line characteristics. This technique holds even for those cases when different lines have non-negligible variations in internal velocity dispersion ($b$) and central optical depth ($\tau_{0}$). Since our COS observations sample many lines of sight within the 2.5\arcsec aperture, we expect a regular distribution of kinematics associated with the single absorbing components, allowing us to utilize the single-velocity component approximation. \par
\begin{figure*}
   	  \centerline{\includegraphics[scale=0.31]{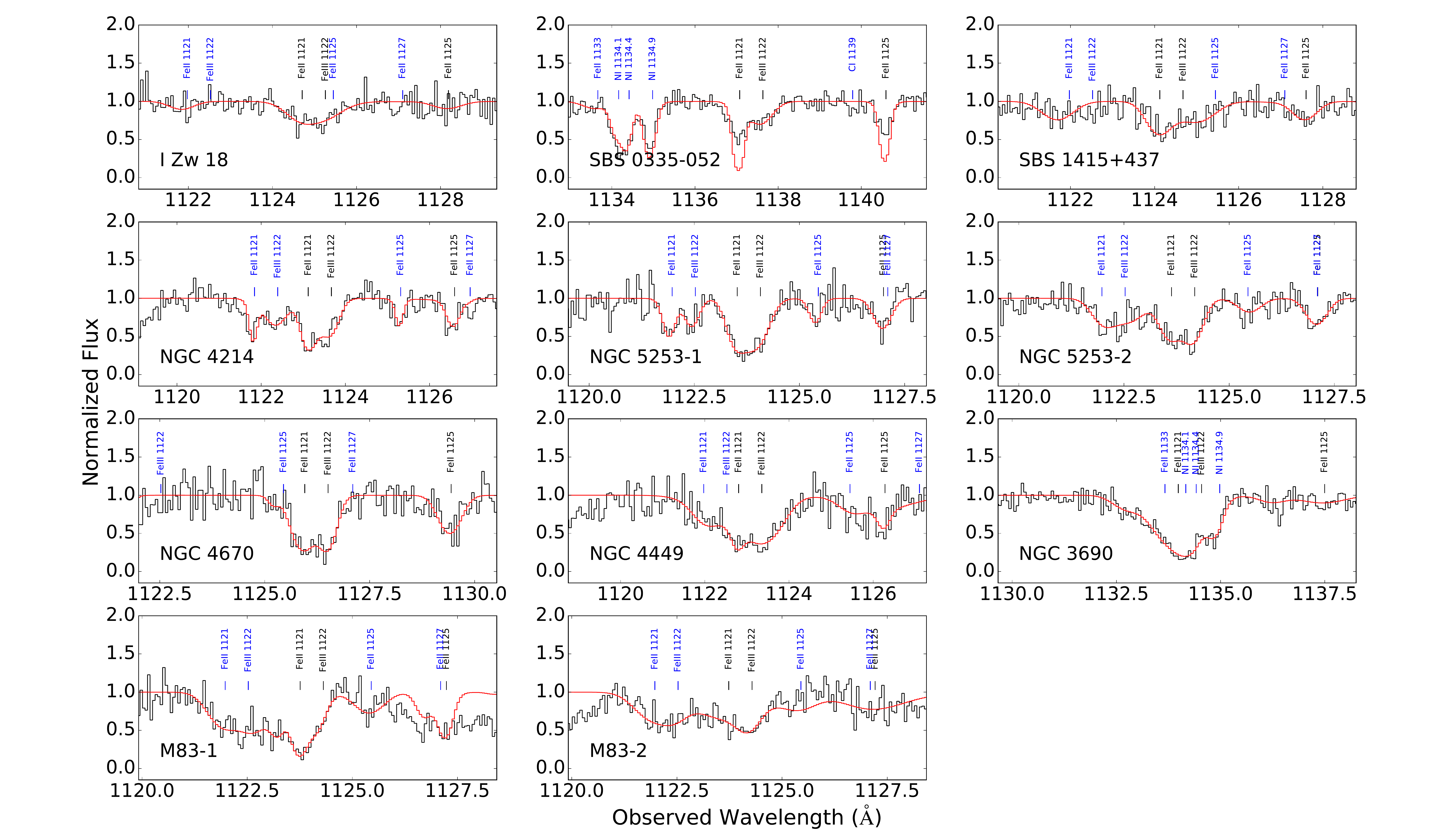}}
      \caption{Absorption-line profiles around the region of the \ion{Fe}{3} line at $\lambda=$1122 \r{A} for the SFG sample studied here. In black solid lines we show the normalized spectroscopic observations. The red solid curves correspond to the best fit profile. The individual ticks above the spectra mark the location of the absorption lines considered in the profile fitting procedure. We display with blue labels those lines corresponding to MW absorption, and in black those intrinsic to the observed galaxy. }
         \label{fig:fe3_fits}
   \end{figure*}

 
In the analysis done as part of this work, we maintain a simple approach when determining the column densities applying the line-profile fitting method. Given the resolution of our COS observations we avoid introducing additional parameters such as velocity distributions and number of intervening clouds. Although in the single-velocity approximation the internal velocity dispersion parameter, $b$, has no precise physical meaning (given that it is the result of combining the multiple line Doppler widths and the many velocity separations between the different line components), the column density is indeed well constrained  \citep{jen86}. To further constrain the column densities measured here we use multiple lines of the same ion with different $f\lambda$ values, where $f$ is the oscillator strength, and $\lambda$ is the rest-frame wavelength of the line of interest. To avoid any saturation issues affecting the strongest lines, the work presented here makes use of only those lines with the weakest $f$ values: \ion{Fe}{3} $\lambda$1122, \ion{Fe}{2} $\lambda$1081, $\lambda$1083, $\lambda$1096, $\lambda$1112, $\lambda$1121, $\lambda$1125, $\lambda$1127, $\lambda$1133, $\lambda$1142, $\lambda$1143, $\lambda$1608, $\lambda$1611 lines. The \ion{Fe}{2} lines used were subject to the target and their rest-wavelength coverage. In Figure \ref{fig:fe3_fits} we show the best fits along with the COS data for each target. We note the strong hidden saturation in SBS 0335-052 clearly shown in the top middle panel. For this target we use the weakest lines when estimating the column densities for \ion{Fe}{2}. \par 

\subsection{COS Line Spread Function}\label{sec:lsf}
As part of the line-profile fitting we take into account the LSF of the COS instrument for the corresponding configuration. The final spectral resolution of the observations is defined by the extension of the source within the 2.5\arcsec$\:$ COS aperture. We derive the final spectral resolution of our data following
\begin{equation}\label{eq:resolution}
{\rm FWHM^{2}_{spec} = \rm FWHM^{2}_{LSF} + FWHM^{2}_{im} - FWHM^{2}_{PSF}}
\end{equation}

where FWHM$\rm_{LSF}$ is the instrument spectral response for a point source in the corresponding COS setting (grating and central wavelength), FWHM$\rm_{im}$ is the extension of the source as measured from the NUV TA images, and FWHM$\rm_{PSF}$ is the FWHM of the point-spread function on the NUV TA images ($\sim$2 pixels). We broadened the intrinsic LSF profile using the $\rm FWHM^{2}_{spec}$ as listed in the last column of Table \ref{tab:obs_log} to accurately account for the source extension. In general we found that the sources in our sample are only slightly extended providing average resolution of $\sim$31 and 30 km s$^{-1}$ for the G130M/1222 and G160M/1623 configurations, respectively. We note that the resolution of these new data differ from those in \citet{jam14} due to the different orientation of the spectroscopic observations and the LP of the instrument.\par

\subsection{Photoionization Modeling}
We estimate the amount of ionized gas along the line of sight of our targets contaminating and contributing to the column densities measured in the neutral gas. To calculate corrections for this effect we simulate the conditions in the interstellar matter using the spectral synthesis code \texttt{CLOUDY} \citep{fer17}. Our approach includes the modeling of the individual star clusters as spherically symmetric gas clouds surrounding a single ionizing source of a given effective temperature and UV luminosity. In Table \ref{tab:tab_cloudy} we list the input parameters used for the \texttt{CLOUDY} ad-hoc models for each of our targets. Our primary goal is to obtain a simplified model of both components, the ionized and neutral gas, along a single line of sight for each galaxy. We note that in reality the true geometry of these targets is more complex than our simplified assumptions. Similar to the work done in \citet{jam14} we generate ionization models covering each of the metallicities ($Z$ = 0.03-- 3.24 $Z_{\sun}$) and neutral hydrogen column densities (log[$N$(\ion{H}{1})/cm$^{-2}$] = 18.44-- 21.70, as measured by \citealt{jam14}) of our SFG sample as listed in Table \ref{tab:tab_cloudy}. For these photoionization models we assume metallicities similar to those from previous studies listed and referenced in Table \ref{tab:tab_cloudy}.  \par
\begin{figure*}
   	  \centerline{\includegraphics[scale=0.3]{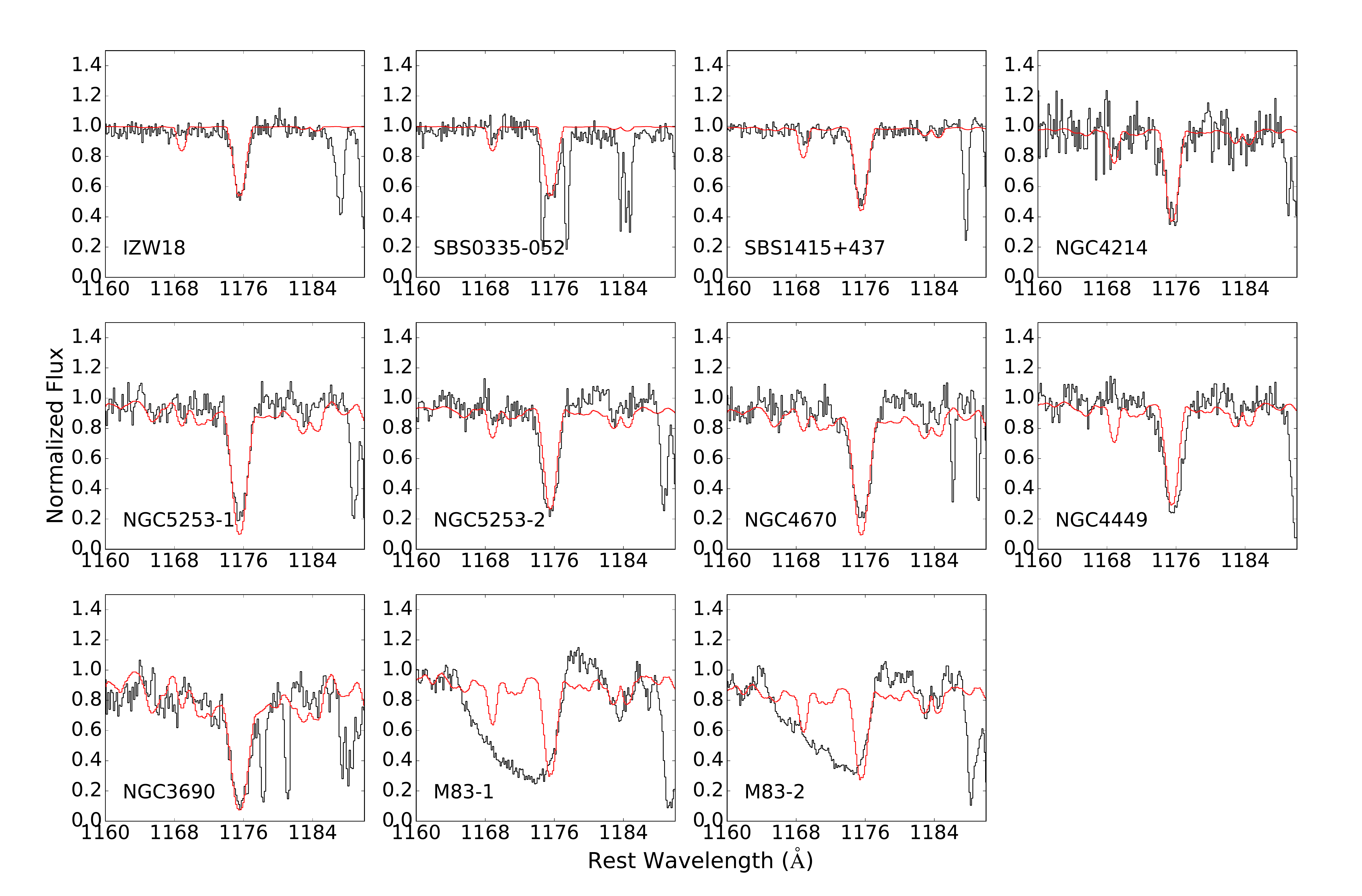}}
      \caption{Absorption-line profiles around the region of the \ion{C}{3} line at $\lambda$ = 1175 \r{A} for the SFG sample studied in this work. The red solid lines correspond to the best fit from a grid of models generated with \texttt{TLUSTY} used to obtain the best effective temperature of the observed stellar population (listed in Table \ref{tab:tab_cloudy}).}
         \label{fig:teff_models}
   \end{figure*}

We approximate the effective temperature, T$_{\rm eff}$, of the observed stellar populations by fitting the COS observations to model stellar spectra generated with \texttt{TLUSTY} \citep{hub95}. We create a grid of stellar models accounting for stellar rotation of $\sim$200 km s$^{-1}$, spanning metallicities of 0.001 $<Z_{\sun}<$ 2 and temperatures of 32,500 $<$T(K)$<$ 55,000. For each target we select only those models that have the same stellar metallicity as their nebular gas  \citep[several studies have shown this to be an accurate assumption, see][]{her17, chi19} but different effective temperature. We focus the fitting procedure on the \ion{C}{3} $\lambda$1175 line, minimizing $\chi^2$ between the rest wavelength range of 1160-1185 \r{A}. This line provides an upper limit in the FUV on the stellar temperature with the strength of the line being inversely proportional to the temperature of the stars \citep{pel02}. In Figure \ref{fig:teff_models} we show the COS observations in black along with the best stellar models in red. For the targets in M83 we identify notable broad blueshifted absorption (typical signature of gas outflow) along with redshifted emission profiles, known as P-Cygni profiles associated with stellar winds in massive stars. The model grid generated with \texttt{TLUSTY} assumes static atmospheres which preclude the inclusion of winds. For all the above mentioned reasons we only fit the red part of the absorption line of these two targets to obtain a rough estimate of the effective temperature of their ionizing sources. We find that the estimated effective temperature has a minimal effect on the relative mixture of neutral and ionized gas within our models translating into negligible effects on our estimated ICFs (see Section \ref{sec:grid_icfs} for more details). We list in Table \ref{tab:tab_cloudy} the adopted T$_{\rm eff}$ values for all the targets in our sample.\par
The UV luminosity of the targets was estimated using the ACS/SBC images. We apply the new sensitivity and aperture corrections from \citet{avi17} and \citet{avi16}, respectively. \par

To obtain accurate estimates of the volume density of the gas for each of the targets we use the \ion{Fe}{3}/\ion{Fe}{2} ratio measured from our COS observations. In the sixth column of Table \ref{tab:tab_cloudy} we list the $\log$[$N$(\ion{Fe}{3})/$N$(\ion{Fe}{2}] ratios measured as part of this work.\par

The photoionization models tailored this time to the physical properties of each galaxy were created using the parameters in Table \ref{tab:tab_cloudy}, and a radius from the central ionizing source of 10 pc \citep{jam14}. As a first step we create a grid of models, one set for each target with a given $Z$ and $N$(\ion{H}{1}), with constant-density gas over a range of volume densities ($-$6 $<$ $\log[n$(H)/ cm$^{-3}$] $<$ 6, in steps of $\log[n$(H)/ cm$^{-3}$] = 0.5). The models are stopped once the desired galaxy \ion{H}{1} column density is reached. We note that by stopping the models at the measured \ion{H}{1} column density we ensure that the models properly account for an adequate amount of neutral gas along the line of sight of each of the galaxies. From the different model grids, one grid for each galaxy, we select the appropriate volume density, $\log[n$(H)/ cm$^{-3}$], using the \ion{Fe}{3}/\ion{Fe}{2} ratio listed in Table \ref{tab:tab_cloudy}. Once we obtain the optimal $\log[n$(H)/ cm$^{-3}$] we use it to generate the best photoionization model which simulates the gas and column densities associated with both the ionized and neutral gas (\ion{H}{1}+\ion{H}{2}). The best volume density values as estimated from our models are listed in the last column of Table \ref{tab:tab_cloudy} \par
Using the \ion{H}{1}+\ion{H}{2} model we determine the location of the photodissociation region (PDR) by identifying  the depth through the cloud at which the ionic fraction of \ion{H}{2} is equal to that of \ion{H}{1}. One last model for each galaxy is created using the same model parameters, however, this last model is stopped at a depth equal to the PDR edge. This last model provides us with estimates of the column density for each element associated with the ionized gas alone, \ion{H}{2}. We use Equation \ref{eq:hi} to assess the column densities of each ion, $X^{i}$ (the dominant ionization stage), arising only from the \ion{H}{1} region,

\begin{equation}\label{eq:hi}
N(X^{i})_{\rm HI} = N(X^{i})_{\rm HI+HII} - N(X^{i})_{\rm HII}
\end{equation}

\noindent ICF$_{\rm ionized}$: for each ion this correction factor accounts for the contaminating ionized gas arising from the \ion{H}{2} region along the line of sight for each of the targets. We estimate ICF$_{\rm ionized}$ by comparing the column densities of the total gas, $N(X^{i})_{\rm HI+HII}$, with the column densities of the neutral gas, $N(X^{i})_{\rm HI}$, alone,
\begin{equation}\label{eq:icf_ionized}
\mathrm{ICF}(X^{i})_{\rm ionized} = \log[N(X^{i})_{HI+HII}]- \log[N(X)_{HI}]
\end{equation}

\noindent ICF$_{\rm neutral}$: for each element this correction factor accounts for the \textit{classical} ionization effects in ISM studies. It originates from the fact that when estimating chemical abundances of the neutral gas, one might be limited by the sampling of absorption lines of the dominant ionization stage of element $X$, while absorption from higher ionization stages are not available but need to be taken into account. This correction is estimated by comparing the column densities of higher ionization states to those of the dominant lower ionization species for a given element within the neutral gas only. We assume that the total column density for each element is given by the following relation

\begin{equation}
N(X) = N(X^{i}) + N(X^{i+1})
\end{equation} 

where $N(X^{i})$ is the column density of the dominant state of ionization in the neutral gas for element $X$ (i.e. \ion{Fe}{2}), and $N(X^{i+1})$ is the column density of the higher ionization state of the same element (i.e. \ion{Fe}{3}). The ICF$_{\rm neutral}$ that is to be applied to the dominant ion, $X^{i}$, in the neutral gas, is then defined as follows

\begin{equation}
\mathrm{ICF}(X^{i})_{\rm neutral} = \log\bigg[\frac{N(X^{i}) + N(X^{i+1})}{N(X^{i})}\bigg]
\end{equation}

We note that the ionization corrections inferred above use the default solar abundance ratios in \texttt{CLOUDY}. However, we want to point out that we do not expect solar abundance ratios in our targets as those are affected by their specific star formation histories. We also highlight that our simulations do not require the inclusion of cosmic rays background as this is only a secondary source of ionization with negligible effects on our gas abundances.

\begin{table*}
\caption{Input parameters for the \texttt{CLOUDY} models tailored to each of the galaxies in our sample}
\label{tab:tab_cloudy}
\centering 
\begin{tabular}{ccccccc}
\hline \hline
Target & $Z$ & T$_{\rm eff}$ & $\log$[L$_{\rm UV}$] & $\log[N($\ion{H}{1}$)^{a}$] &  $\log$[$N$(\ion{Fe}{3})/$N$(\ion{Fe}{2})] & log[$n$(H)]\\
 & ($Z_{\sun}$) &(K) & (erg s$^{-1}$) &  (cm$^{-2}$) & & (cm$^{-3}$) \\
\hline\\
        I Zw 18& 0.03 & 40,000&41.34 &21.28 & $-$0.194 $\pm$ 0.164 & 3.03 \\	 	 
	SBS 0335-052&0.04 &40,000	 &42.60 &21.70 &$-$1.109 $\pm$ 0.089 & 5.50\\ 	 	 	  	 
	SBS 1415+437& 0.08&40,000&41.09 &21.09 & $-$0.615 $\pm$ 0.253 & 3.73\\	 	 	  	  	 	 
	NGC 4214& 0.32& 40,000&40.86 &21.12 & $-$0.291 $\pm$ 0.427 & 2.77\\ 	
	NGC 5253-1 &0.32 & 32,500&40.52 &21.20 & $-$0.213 $\pm$ 0.118 & 2.27 \\
	NGC 5253-2 & 0.32 & 37,500&40.27 &20.65 & $-$0.384 $\pm$ 0.108 & 3.04\\ 	  	 
	NGC 4670& 0.32 &32,500& 42.23&21.07 & $-$0.534 $\pm$ 0.103 & 4.52\\
	NGC 4449& 0.41 & 40,000 &40.31 &21.14 & $-$0.496 $\pm$ 0.364& 2.45\\	 	 	 		 	  
	NGC 3690& 1.29 &32,500&42.24 &20.68$^b$ & $+$0.120 $\pm$ 0.371& 4.17\\ 		 	 	 	 	 
	M83-1& 3.24 &42,500& 40.94&19.60 & $-$0.798 $\pm$ 0.395 & 4.68\\	 	 	 	 	 
	M83-2& 3.24& 42,500 & 41.40&18.44 & $-$0.824 $\pm$ 0.213 & 5.72\\	 	 	 	 	 
\hline
\end{tabular}
\begin{minipage}{15cm}~\\
\textsuperscript{$a$}{Published in \citet{jam14}.}\\
\textsuperscript{$b$}{Total hydrogen column density for both velocity components as observed in \citet{jam14}.}\\
 \end{minipage}
\end{table*}

\section{Discussion}\label{sec:discussion}

\subsection{Ionization Corrections from ad-hoc models}\label{sec:sfgs_icfs}
The final ICF$_{\rm ionized}$ and ICF$_{\rm neutral}$ to be applied to the dominant ions to obtain the total column densities of a given element in our sample are shown in Table \ref{tab:icfs}. We also include in this table the total ionization correction factors (ICF$_{\rm TOTAL}$) applied to the measured column densities as published by \citet{jam14}. We point out that these ICFs are tailored for the galaxies studied here which span a wide range of metallicities, hydrogen column densities, luminosities, and ionizing temperature (temperature of the ionizing source as listed in Table \ref{tab:tab_cloudy}). Additionally, we list the updated interstellar element abundances for our sample of SFGs in Table \ref{tab:tab_abun} and provide a one-on-one comparison between the final abundances using the ionization corrections estimated as part of this work (new) and those from \citet[][old]{jam14} in Figure \ref{fig:old_new}.\par
As mentioned before \citet{jam14} obtained ICFs for the same SFG sample studied here. However, that work was limited by the wavelength range of the original COS observations, preventing us from calculating a \ion{Fe}{3}/\ion{Fe}{2} ratio for each of the galaxies. \citet{jam14} apply a more conservative strategy, compared to the work presented here, since the grid of models was generated using the average luminosity of the galaxies (40.05 erg s$^{-1}$, as measured from the counts within the COS aperture on the ACS/SBC images), as well as the average effective temperature of their ionizing sources (36,000 K). We note that since the time of the publication of \citet{jam14}, the ACS team has released new zeropoints for the SBC filters and new time dependent sensitivity \citep[TDS; ][]{avi17} curves. The combination of these two effects is the main cause for the differences between the luminosities used in \citet{jam14} and those estimated in this work. \par

\subsection*{ICF$_{\rm ionized}$}
Similar to the findings in \citet{jam14} we identify an overall positive correlation between the ICF$_{\rm ionized}$ and the metallicity of the sources for each of the elements (see Table \ref{tab:icfs}). Our ICF$_{\rm ionized}$ tend to increase, on an element base, towards higher metallicities. We note that the here inferred ICF$_{\rm ionized}$ values are comparable to those estimated in the work by \citet{jam14} for most targets with the exception of NGC3690 and M83-2. \par
For NGC3690, \citet{jam14} obtained two different sets of  ICF$_{\rm ionized}$, one for each of the individual velocity components identified in the spectrum for this target. For the work detailed here, we instead combine both components into a single one since our \ion{Fe}{3} analysis is not able to resolve both of the components observed in this line of sight. We then combine their measurements of the neutral hydrogen column densities into a single value, and obtained  ICFs for the combined components. We also point out that in contrast to the results in \citet{jam14}, the tailored approach applied here allowed us to extract ICF$_{\rm ionized}$ estimates for the most extreme environment in our sample, M83-2, being the most metal rich target with the lowest $N$(\ion{H}{1}) value in our sample. For this target we observed the highest ICF$_{\rm ionized}$ ($\sim$0.2-0.7 dex) across all elements.\par

\subsection*{ICF$_{\rm neutral}$}
We compare the ionization correction factors for the neutral gas, ICF$_{\rm neutral}$, inferred here to those from \citet{jam14}. There is a general agreement between the two studies where for most of the elements and targets the corrections for the neutral gas are negligible with two exceptions, the M83-2 target, and the overall corrections for nitrogen. \par
Similar to the ICF$_{\rm ionized}$ estimates, \citet{jam14} are not able to calculate ICF$_{\rm neutral}$ for the most metal rich target, M83-2. With the new approach described here, we are able to obtain ionization corrections for this target. Comparing the corrections for each element we find that the ICF$_{\rm neutral}$ values for M83-2 have the highest corrections, amongst the rest of the galaxies, especially for H, N, and O. These higher-than-average corrections could be attributed to the extreme metallicity environment observed in M83-2. \par
\begin{figure}
   	  \centerline{\includegraphics[scale=0.43]{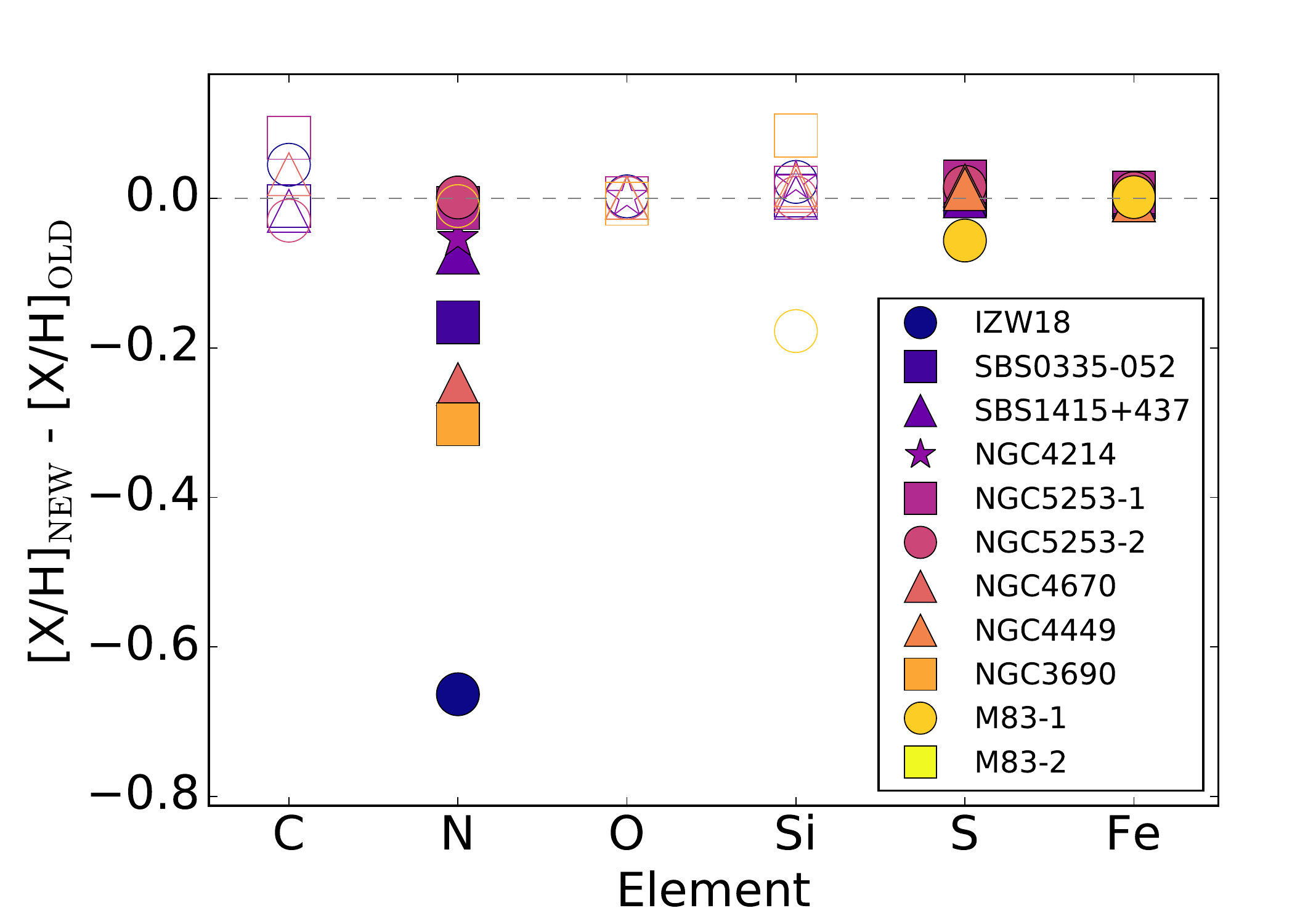}}
      \caption{Difference between the corrected element abundances, using the new ICFs inferred here, and those from \citet{jam14} as a function of element. The different symbols represent the different galaxy measurements. Empty (solid) symbols correspond to abundances that are (are not) lower limits as detailed in \citet{jam14}.}
         \label{fig:old_new}
   \end{figure}

A one-on-one comparison between the ICF$_{\rm neutral}$ values inferred here and those published by \citet{jam14} shows that overall there are no major differences, except for nitrogen, where the ICF$_{\rm neutral}$ values are remarkably higher in this new work than previously considered. This trend is clearly displayed in Figure \ref{fig:old_new} where the differences between the final abundances using the total ionization corrections estimated as part of this work (new) and those from \citet[][old]{jam14} lie in the applied ICFs. The largest deviation in nitrogen is found in I Zw 18 (dark blue circle in Figure \ref{fig:old_new}). \par
\begin{figure*}
   	  \centerline{\includegraphics[scale=0.44]{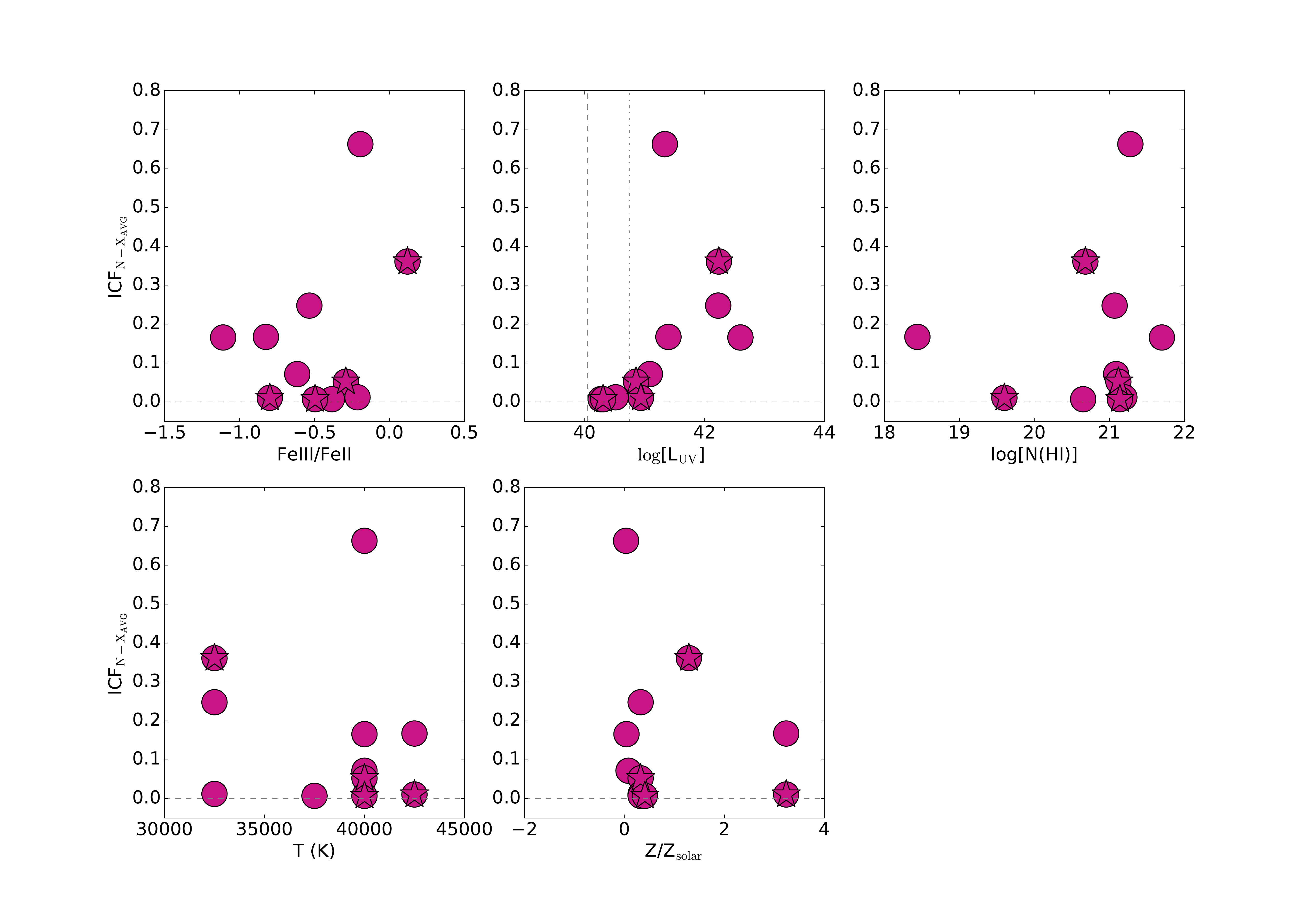}}
      \caption{Difference between the ICF$_{\rm neutral}$ for nitrogen and the average for all the elements as a function of the different physical parameters: \ion{Fe}{3}/\ion{Fe}{2}, $\log$[L$_{\rm UV}$], $\log$[N(\ion{H}{1})], effective temperature, and metallicity. We show in stars those targets with two velocity components, but their \ion{Fe}{3} lines appeared unresolved. To help guide the eye we also show horizontal dashed lines at ICF$_{\rm N-X_{\rm AVG}}$ = 0 in each of the panels. No trends are identified except for the top middle panel where it is clear that the nitrogen correction is substantially larger than that of the other elements at $\log$[L$_{\rm UV}$] $>$ 41 erg s$^{-1}$  suggesting a possible correlation between luminosity and the ICF$_{\rm neutral}$ values. The vertical dashed line in that panel shows the luminosity value used in the work of \citet{jam14}, $\log$[L$_{\rm UV}$] = 40.05 erg s$^{-1}$. The vertical dotted dashed line shows the luminosity value at which the ICF$_{\rm neutral}$ for nitrogen begins to deviate from that of the other elements ($\log$[L$_{\rm UV}$] = 40.75 erg s$^{-1}$). }
         \label{fig:lum}
   \end{figure*}

Given that the ICF$_{\rm neutral}$ estimates for nitrogen appear to be consistently higher than the average value of the rest of the elements ($\sim$0.004) we compare in Figure \ref{fig:lum} the ICF$_{\rm neutral}$ values for nitrogen against the average corrections for each element  (ICF$_{\rm N-X_{\rm AVG}}$) as a function of the different input model parameters, $\log$[$N$(\ion{Fe}{3})/$N$(\ion{Fe}{2})], $\log$[L$_{\rm UV}$], $\log N$(\ion{H}{1}), T$_{\rm eff}$ and $Z$. From this figure we can see that there are no clear trends for most of the parameters, with the exception of the luminosity. The top middle panel shows that the ICF$_{\rm neutral}$ for nitrogen is comparable to the one for the other elements until $\log$[L$_{\rm UV}$] $\sim$ 40.75 erg s$^{-1}$ is achieved (shown with a vertical dotted dashed line) at which point the correction for nitrogen begins to become substantially larger than the rest of the elements.
Since in the work of \citet{jam14} a luminosity of $\log$[L$_{\rm UV}$] = 40.05 erg s$^{-1}$  (shown in the top middle panel of Figure \ref{fig:lum} as a dashed vertical line) was used to calculate the ICF values for all of the targets, such high ICF$_{\rm neutral}$ values for nitrogen had not been previously derived. 


In general, the ICF$_{\rm neutral}$ aims to correct for the amount of gas from a higher ionization stage for a given element, in this case \ion{N}{2}, present in the \ion{H}{1} region. Given the higher-than-average ICF$_{\rm neutral}$ values for nitrogen at $\log$[L$_{\rm UV}$] $\gtrsim$ 40.75 erg s$^{-1}$ we can safely assume that there is a non-negligible amount of  \ion{N}{2} in the neutral gas of some of the galaxies studied as part of this work. The possible correlation between the nitrogen ICFs and the luminosity of the source shown in Figure \ref{fig:lum} leads us to conclude that the higher photon energies emanating from the ionizing sources are able to ionize nitrogen at larger distances than the rest of the elements. \par

\begin{table*}
\caption{Ionization Correction Factors for the SFGs in this work}
\label{tab:icfs}
\centering 
\begin{tabular}{cccccccccc}
\hline \hline
\multicolumn{10}{c}{ICF$_{\rm ionized}$}\\
 \hline
Target & \ion{H}{1} & \ion{C}{2} & \ion{N}{1} & \ion{O}{1} & \ion{Si}{2} & \ion{P}{2} & \ion{S}{2}& \ion{Fe}{2}& \ion{Ni}{2}\\
\hline\
        I Zw 18& 0.002 & 0.099 & 0.001 &0.002 & 0.034 & 0.090 & 0.023 & 0.006 & 0.013  \\	 	 
	SBS 0335-052 & 0.000 & 0.010 & 0.000 & 0.000 & 0.004 & 0.013 & 0.005 & 0.001 & 0.004\\ 	 	 	  	 
	SBS 1415+437& 0.002 & 0.055 & 0.000 & 0.002 & 0.022 & 0.066 &0.024 & 0.006 & 0.017 \\	
	NGC 4214& 0.002 & 0.098 & 0.000 & 0.002 & 0.044 & 0.081 & 0.027 &  0.009 & 0.016\\ 	
	NGC 5253-1 & 0.002 & 0.132 & 0.001 & 0.002 & 0.037 & 0.075 & 0.034 & 0.009 & 0.014 \\
	NGC 5253-2 & 0.004 & 0.113 & 0.001 & 0.004 & 0.064 & 0.116 & 0.059 &  0.020 & 0.040  \\ 	  	 
	NGC 4670& 0.002 & 0.104 & 0.001 & 0.002 & 0.031 & 0.079 & 0.038 & 0.009 & 0.018   \\
	NGC 4449& 0.002 & 0.072 & 0.000 & 0.002 & 0.039 & 0.066 & 0.024 & 0.009 & 0.014 \\	 	 	 		 	  
	NGC 3690& 0.005 & 0.298 & 0.002 & 0.005 & 0.126 & 0.191 & 0.097 & 0.036 & 0.037  \\ 		 	 	 	 	 
	M83-1& 0.021 & 0.183 & 0.014 & 0.024 & 0.213 & 0.204 & 0.156 & 0.163 & 0.168   \\	 	 	 	 	 
	M83-2& 0.274 & 0.687 & 0.231 & 0.295 & 0.707 & 0.703 & 0.668 & 0.661 & 0.703 \\	 	 	 	 	 
 \hline
\multicolumn{10}{c}{ICF$_{\rm neutral}$}\\
 \hline
Target & \ion{H}{2} & \ion{C}{3} & \ion{N}{2} & \ion{O}{2} & \ion{Si}{3} & \ion{P}{3} & \ion{S}{3}& \ion{Fe}{3}& \ion{Ni}{3}\\
\hline\
        I Zw 18& 0.003 & 0.005 & 0.665 & 0.000 & 0.002 & 0.002 & 0.002 & 0.000 & 0.000\\	 	 
	SBS 0335-052 & 0.000 & 0.000 & 0.166 & 0.000 & 0.000 & 0.000 & 0.000 & 0.000 & 0.000  \\ 	 	 	  	 
	SBS 1415+437& 0.001 & 0.001 & 0.072 & 0.000 & 0.000 & 0.001 & 0.001 & 0.000 & 0.000\\	
	NGC 4214& 0.001 & 0.002 & 0.053 & 0.000 & 0.001 & 0.001 & 0.001 & 0.000 & 0.000  \\ 	
	NGC 5253-1 & 0.001 & 0.001 & 0.013 & 0.001 & 0.002 & 0.001 & 0.001 & 0.001 & 0.001 \\
	NGC 5253-2 & 0.001 & 0.001 &  0.008 & 0.001 & 0.001 & 0.001 & 0.001 & 0.001 & 0.001  \\ 	  	 
	NGC 4670& 0.001 & 0.001 & 0.248 & 0.000 &  0.000 &  0.000 &  0.000 &  0.000 & 0.000 \\
	NGC 4449& 0.001 & 0.001 & 0.008 & 0.000 & 0.001 & 0.001 & 0.001 & 0.000 & 0.000  \\	 	 	 		 	  
	NGC 3690& 0.004 & 0.007 & 0.364 & 0.002 & 0.002 & 0.003 &0.003 & 0.000& 0.002 \\ 		 	 	 	 	 
	M83-1& 0.010 & 0.001 & 0.013 & 0.008 & 0.000 &  0.000 & 0.002 & 0.000 & 0.000  \\	 	 	 	 	 
	M83-2& 0.168 & 0.006 & 0.209 & 0.136 & 0.001 & 0.004 & 0.017 & 0.003 & 0.000  \\
  \hline
\multicolumn{10}{c}{ICF$_{\rm TOTAL}$}\\
 \hline
Target & H & C & N & O & Si & P & S & Fe & Ni \\
\hline\
        I Zw 18& 0.000 & 0.095 & $-$0.664 & 0.002 & 0.032 & 0.088 & 0.021 & 0.006 & 0.013\\	 	 
	SBS 0335-052 & 0.000 & 0.010 & $-$0.166 &  0.000 & 0.004 & 0.013 &  0.005 & 0.001 &0.004 \\ 	 	 	  	 
	SBS 1415+437& 0.001 & 0.054 & $-$0.072 &  0.001 & 0.022 & 0.066 & 0.023 & 0.006 & 0.017 \\	
	NGC 4214& 0.001 &  0.096 & $-$0.053 &0.002 &  0.043 & 0.080 & 0.026 & 0.009 & 0.016 \\ 	
	NGC 5253-1 & 0.001 & 0.132 & $-$0.012 & 0.001 & 0.035 & 0.074 &  0.033 & 0.009 & 0.013\\
	NGC 5253-2 &  0.002 & 0.112 & $-$0.007 &0.003 &  0.063 & 0.115 & 0.058 & 0.019 & 0.039 \\ 	  	 
	NGC 4670& 0.001 & 0.103 & $-$0.248 &  0.001 & 0.031 & 0.079 & 0.038 &0.008 & 0.018  \\
	NGC 4449& 0.001 & 0.071 & $-$0.007 & 0.002 & 0.039 & 0.064 &  0.023 & 0.008 &  0.014 \\	 	 	 		 	  
	NGC 3690& 0.001 & 0.291 & $-$0.362 &  0.003 & 0.125 & 0.188 & 0.093 & 0.036 &  0.036 \\ 		 	 	 	 	 
	M83-1&  0.011 & 0.183 & 0.000 & 0.016 & 0.213 & 0.203 & 0.154 & 0.162 & 0.167 \\	 	 	 	 	 
	M83-2& 0.106 & 0.680 & 0.022 & 0.159 &  0.706 & 0.699 & 0.651 & 0.658 & 0.702 \\
 \hline
\end{tabular}
\end{table*}

\begin{table*}
\caption{Updated Interstellar Abundances}
\label{tab:tab_abun}
\centering 
\begin{tabular}{ccccccccc}
\hline \hline
 & \multicolumn{8}{c}{[X/H]$_{\rm ICF}$}\\
 \hline
Target& C &  N& O& Si& P& S& Fe&  Ni\\
\hline\\
     I Zw 18& $>-$3.39 & $-$2.07 & $>-$3.17 & $>-$2.46 & $--$ & $-$1.69 & $-$2.26 & $--$ \\	 	 
     SBS 0335-052& $>-$3.73 & $-$2.53 & $>-$3.62 & $>-$2.83 & $-$2.15 & $-$1.87 & $-$1.88 & $-$1.96\\ 	 	 	  	 
	SBS 1415+437& $>-$3.06 & $-$2.28 & $>-$2.91 & $>-$2.20 & $-$1.14 & $-$1.27 & $>-$1.77 & $-$2.07 \\	 	 	  
	NGC 4214& $--$  & $-$1.70 & $>-$2.56 & $>-$1.80 & $-$0.85& $-$0.71 & $-$1.83 & $-$1.88\\ 	
	NGC 5253-1& $>-$2.72 & $-$1.78 & $>-$2.45 & $>-$1.95 & $-$0.94 & $-$0.92 & $-$1.04 & $-$1.75\\
	NGC 5253-2& $>-$2.27 &  $-$1.49 & $>-$2.03 &  $>-$1.41 & $-$0.52 & $-$0.46 & $-$0.50 & $-$1.14\\ 	  	 
	NGC 4670& $>-$2.58& $-$1.23 & $>-$2.50& $>-$1.64& $-$0.74& $>-$0.63& $-$1.51& $-$1.55 \\
	NGC 4449& $--$ & $--$ & $>-$2.29& $>-$1.44& $-$0.50& $-$0.67& $>-$1.61& $--$\\	
	NGC 3690$^{a}$ & $--$& $-$1.03& $>-$1.80& $>-$1.15 &$--$ & $--$ & $-$1.35 & $--$\\
	M83-1 & $--$ & $>-$0.13& $--$ & $>-$0.26& $+$0.88 & $+$0.74 & $>-$0.42 & $-$0.09\\
	M83-2 & $>-$0.35 & $+$0.71& $--$ & $>+$0.14 & $+$1.29 & $>+$1.29 & $+$0.22 & $--$\\	 	 		 	  
 	 	 	 	 
\hline
\end{tabular}
\begin{minipage}{15cm}~\\
\textsuperscript{$a$}{Total Abundances for both velocity components.}\\

 \end{minipage}
\end{table*}

\subsection{The case of I Zw 18}
As detailed in Table \ref{tab:icfs} the ICF$_{\rm neutral}$ for nitrogen required for the SFG I Zw 18 is the highest in our galaxy sample reaching a value of $\sim-$0.7 dex. Such high correction had not been previously derived. This is the reason why in Figure \ref{fig:old_new}  the difference between the abundances using our new ICFs and those from \citet{jam14} is the largest for I Zw 18. This galaxy in particular has some extreme properties. I Zw 18 is the most metal-poor galaxy in our sample, $Z= 0.03 Z_{\odot}$. It also has a relatively high luminosity, $\log$[L$_{\rm UV}$] = 41.34 erg s$^{-1}$, above the threshold discussed in Section \ref{sec:sfgs_icfs}. As we will discuss later in Section \ref{sec:grid_icfs}, from the analysis of Figure \ref{fig:icfs_neutral_grid} showing the ICF$_{\rm neutral}$ values calculated from a uniform model grid, we see that both of these parameters have a strong effect on the ICF$_{\rm neutral}$ for nitrogen. However, considering that \citet{jam14} modeled the I Zw 18 environment with \texttt{CLOUDY} using the low metallicity of this system and did not find any higher-than-normal ICFneutral values for nitrogen, we conclude that the enhanced values for the nitrogen correction are mainly driven by the higher luminosity used in this new work. The large ICF values derived in this particular environment clearly demonstrate that caution must be taken in future studies of galaxies with relatively extreme properties. Determining accurate ionization correction factors will be critical in assessing the true chemical composition of such systems. This is particularly true for high redshift galaxies that are on average more metal poor than present-day galaxies. 

\subsection{Updated abundances}
In Table \ref{tab:tab_abun} we list the new chemical abundances as measured from \citet{jam14} and after applying the new ICF values listed in Table \ref{tab:icfs}. This table, as well as the previously discussed Figure \ref{fig:old_new}, show the immediate effects of our new ionization corrections on the derived abundances (by number relative to those in the sun).  Note that the values for C, O and Si are lower limits due to saturation effects described in \citet{jam14}. \par
Assuming 12+$\log$(O/H)$_{\odot}$ = 8.76  
\citep{caf09} we now compare the oxygen abundances obtained for the neutral gas with those for the ionized gas for a couple of targets in our sample. From Table \ref{tab:tab_abun}, for the most metal poor target, I Zw 18, we estimate a neutral ISM metallicity of 12+$\log$(O/H) $>$ 5.59, consistent with 12+$\log$(O/H) = 7.2 measured in \ion{H}{2} studies \citep{izo99} and listed in Table \ref{tab:gal_prop} as our neutral gas abundance is only a lower limit and not well constrained. \par
Several proxies for oxygen such as \ion{P}{2} and \ion{S}{2} have been proposed and used \citep{jam18} to indirectly derive oxygen abundances when the available O absorption lines suffer from hidden saturation, as is the case for our targets. 
Using the solar ratio of $\log$(S/O)$_{\odot}=-$1.57 by \citet{jam18} to derive  [O/H] for I Zw 18, we instead obtain 12+$\log$(O/H)= 7.0, a value in better agreement with the \ion{H}{2} region study.
The newly calculated oxygen abundance gives a $\log$(N/O) ratio of $-$1.21, in agreement to what is typically observed in metal-poor dwarfs, $\log$(N/O)$\sim-$1.4 \citep[][]{ski13,ann19}. Additionally, we use \ion{S}{2} to indirectly estimate an oxygen abundance for NGC 4449 of 12+$\log$(O/H) = 8.02, similar to what is observed in the ionized gas \citep{ann17}.

\subsection{Ionization corrections for a uniform grid}\label{sec:grid_icfs}
To obtain a better sense of the dependance of ICFs on the physical properties of a galaxy environment, in addition to the tailored ICFs for each of the galaxies in our sample, we have also calculated ICFs from a grid of models covering a wide range of $Z$, T$_{\rm eff}$, $\log$[L$_{\rm UV}$], $N$(\ion{H}{1}), and $\log$[$N$(\ion{Fe}{3}/\ion{Fe}{2})]. The ionization correction factors obtained from the model grid can be used to apply accurate corrections to the neutral gas abundances estimated from a wide variety of galactic environments. We highlight the fact that the galactic environments explored in this grid encompass those typically seen in extragalactic studies of star-forming galaxies and \ion{H}{2} regions. One of the main drivers for estimating these ICFs for such a wide range of environments is to provide the community with the most accurate ICF values applicable to their individual studies. The individual ICF (ionized and neutral) calculated from this uniform grid of models are detailed in the appendix in Tables \ref{tab:icfs_val_grid} and \ref{tab:icfs_val_grid_neutral}. The model parameters explored in this part of the analysis are listed in Table ~\ref{tab:icfs_grid}. We show in boldface the adopted standard model which was defined by the average properties in our SFG sample. We proceed by generating individual photoionization models setting the standard parameters fixed and varying one physical parameter at a time using the values listed in Table ~\ref{tab:icfs_grid}.  

\begin{table*}
\caption{Parameter space explored in the \texttt{CLOUDY} uniform grid}
\label{tab:icfs_grid}
\centering 
\begin{tabular}{ccccc}
\hline \hline
$Z$ & T$_{\rm eff}$ & $\log$[L$_{\rm UV}$] & $\log[N($\ion{H}{1}$)$] &  $\log$[$N$(\ion{Fe}{3})/$N$(\ion{Fe}{2})]\\
($Z_{\sun}$) &(K) & (erg s$^{-1}$) &  (cm$^{-2}$) & \\
\hline\\
         0.03 & 32,500&40.00 &18.4& $-$1.00\\	 	 
	0.10 &35,000	 &40.75 &19.0 &$-$0.50\\ 	 	 	  	 
	0.50&37,500& \textbf{41.25} &20.0 & \textbf{$-$0.49}\\	 	 	  	  	 	 
	0.75& \textbf{38,000}&41.50 &\textbf{20.72} & $-$0.10\\	
	\textbf{0.87} & 40,000&42.25 &21.00& $-$0.05 \\
	1.00 & 42,500 & 43.00 & 22.00 & $+$0.10 \\ 	  	 
	1.25 &--& --&-- & $+$0.50\\
	3.00 &--& --&-- & --\\	 	 	 	 	 
\hline
\end{tabular}
\begin{minipage}{15cm}~\\
\textit{Note:} We show in boldface the parameters used for the adopted standard model.
 \end{minipage}
\end{table*}

\subsection*{ICF$_{\rm ionized}$}
\begin{figure*}
   	  \centerline{\includegraphics[scale=0.38
   	  ]{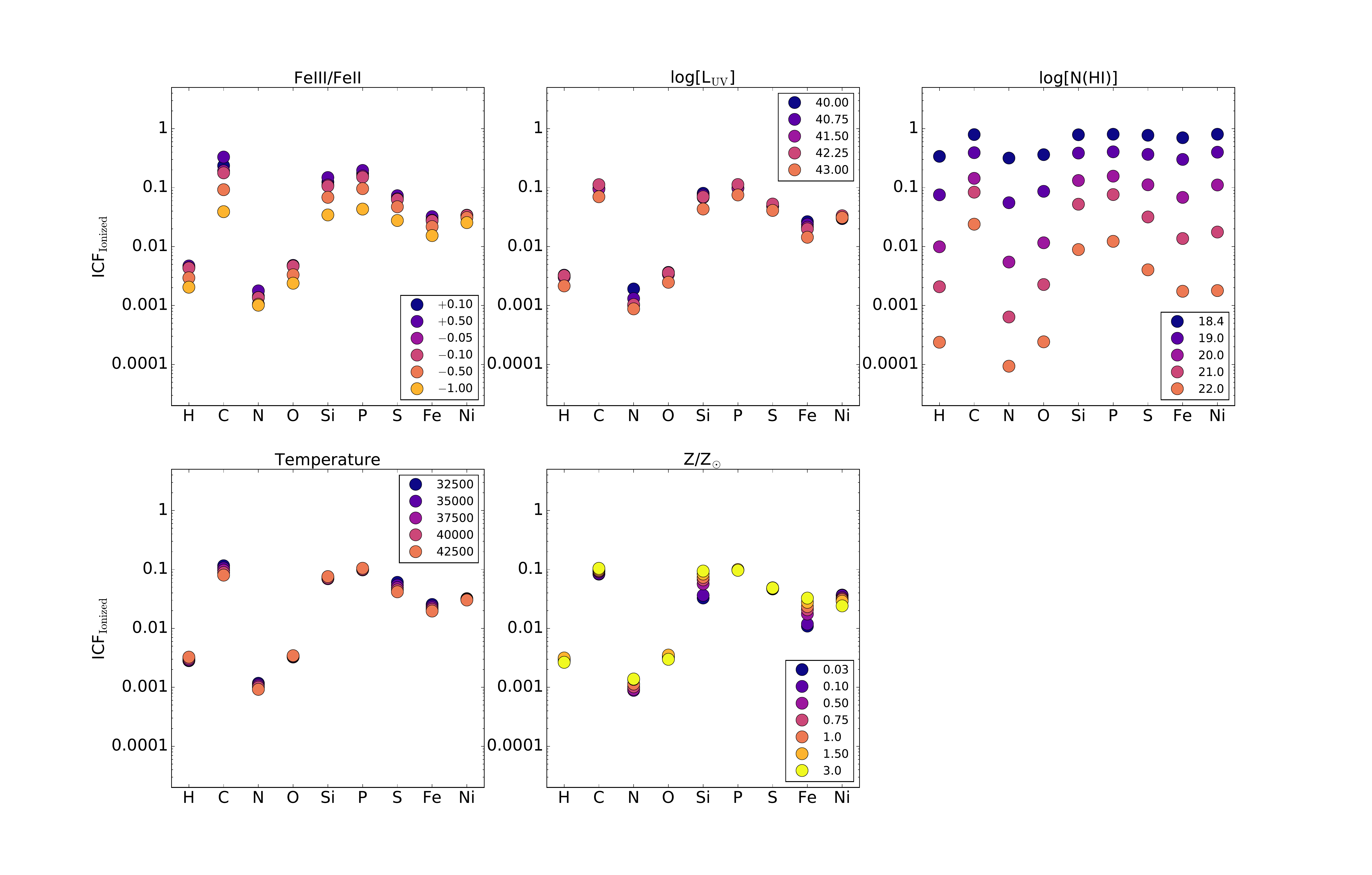}}
      \caption{ICF$_{\rm ionized}$ as a function of element. In each panel we show the corrections estimated using the adopted standard parameters (listed in bold font in Table ~\ref{tab:icfs_grid}) and varying a single property as listed in the title and legend of the subplots.}
         \label{fig:icfs_ionized_grid}
   \end{figure*}

In Figure~\ref{fig:icfs_ionized_grid} we show the variations in the ICF$_{\rm ionized}$ values obtained from the model grid described above. For every model shown in Figure ~\ref{fig:icfs_ionized_grid} we use the adopted standard model properties, and vary a single parameter at a time (shown in the panel title with the specific values shown in the legend of each subplot). Overall we see that the ionization correction factors for contaminating ionized gas, ICF$_{\rm ionized}$, do not vary significantly for each element when changing the individual model properties. This is clearly shown in the Temperature subpanel where varying the effective temperature of the ionizing source has almost a negligible effect on the ICF$_{\rm ionized}$ for each element. The exception to this trend is the column density of \ion{H}{1}. In the top right panel of Figure ~\ref{fig:icfs_ionized_grid} we can see that the highest ICF$_{\rm ionized}$ values are obtained for those models with the lowest $N$(\ion{H}{1}). For such low column densities elements like C and Ni reach ionization corrections as high as 0.8 dex. 

\subsection*{ICF$_{\rm neutral}$}
\begin{figure*}
   	  \centerline{\includegraphics[scale=0.38]{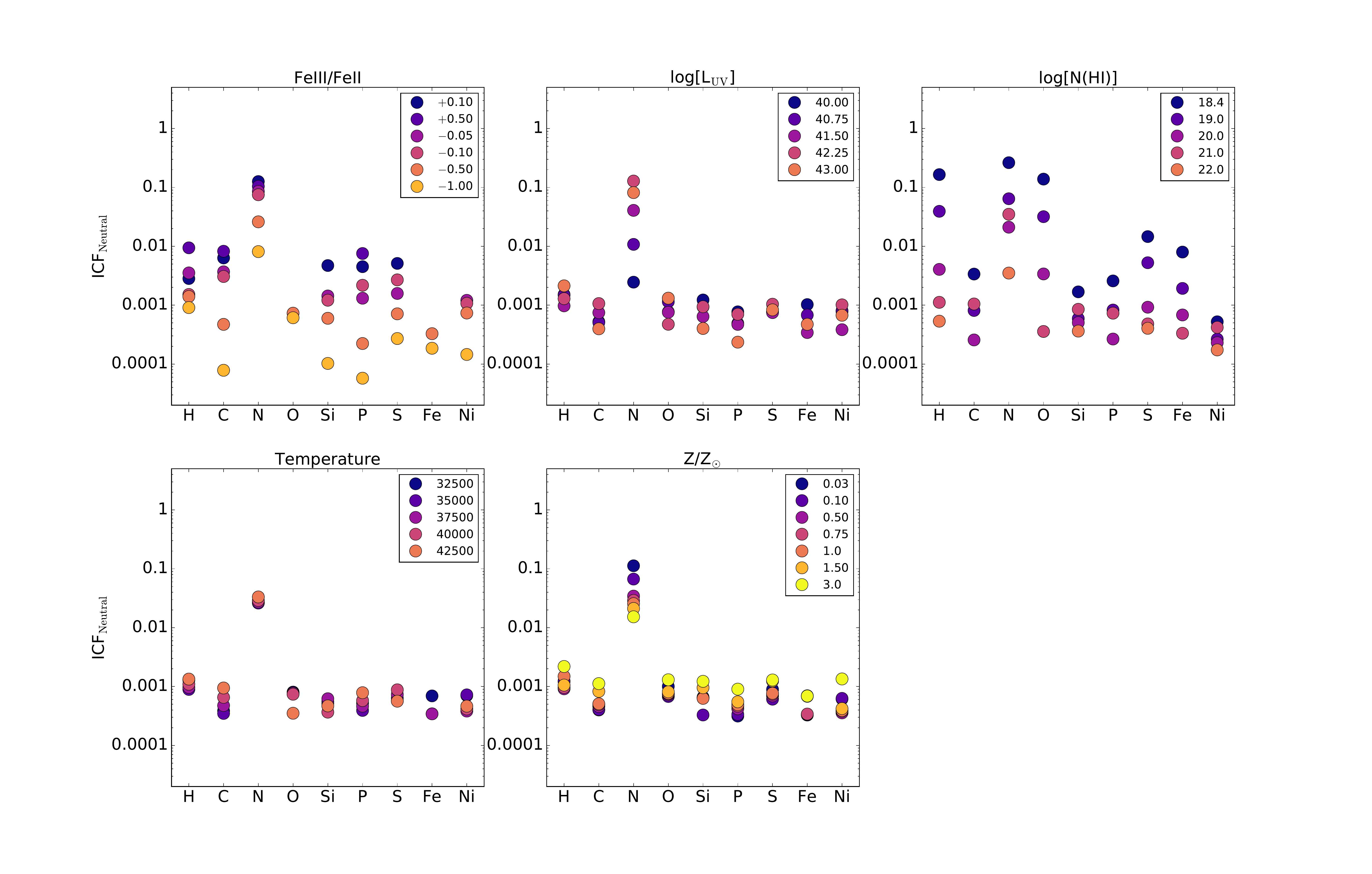}}
      \caption{ICF$_{\rm neutral}$ as a function of element. In each panel we show the corrections estimated using the adopted standard parameters (listed in bold font in Table ~\ref{tab:icfs_grid}) and varying a single property as listed in the title and legend of the subplots.}
         \label{fig:icfs_neutral_grid}
   \end{figure*}

In Figure~\ref{fig:icfs_neutral_grid} we show the ICF$_{\rm neutral}$ values calculated after varying the different parameters in the model grid. Looking at the different panels in Figure ~\ref{fig:icfs_neutral_grid} we can see that there is an overall trend where the neutral corrections for nitrogen are consistently higher than those for the other elements. We point out that in the adopted standard model, we use a $\log$[L$_{\rm UV}$] = 41.25 erg s$^{-1}$, a luminosity in the regime where nitrogen is expected to have higher correction values than the rest of the elements, which could be the main driver in the high ICF$_{\rm neutral}$ values for this element. The proposed  ICF$_{\rm neutral}$ dependance on the luminosity of the ionizing source can be observed in detail in the top middle panel of Figure ~\ref{fig:icfs_neutral_grid} where we show the variations in the ICF$_{\rm neutral}$ values arising when we set the luminosity to a different value. This panel shows that for the model using a $\log$[L$_{\rm UV}$] = 40.00 erg s$^{-1}$, the neutral correction factors are approximately the same for all of the elements (dark blue circles). It is when we increase the luminosity of the ionizing source to $\log$[L$_{\rm UV}$] = 40.75 erg s$^{-1}$ that we begin to see the ICF$_{\rm neutral}$ values for nitrogen deviate from those of the rest of the elements. And as expected and discussed above, higher luminosities than 40.75 erg s$^{-1}$ result in higher ICF$_{\rm neutral}$ estimates for nitrogen.\par
Furthermore, we find that similar to the ICF$_{\rm ionized}$ in general varying the effective temperature of the ionizing source has only a minimal affect on the ICF$_{\rm neutral}$ when looking at individual elements (bottom left panel in Figure ~\ref{fig:icfs_neutral_grid} ). The largest scatter is seen in carbon where the ICF$_{\rm neutral}$ estimates vary negligibly between 0.0004 and 0.0009 dex. We find that the opposite is true for the \ion{H}{1} column density. Our models show that varying $N$(\ion{H}{1}) strongly modifies the ICF$_{\rm neutral}$ values for the neutral gas. \par
\begin{figure*}
   	  \centerline{\includegraphics[scale=0.38]{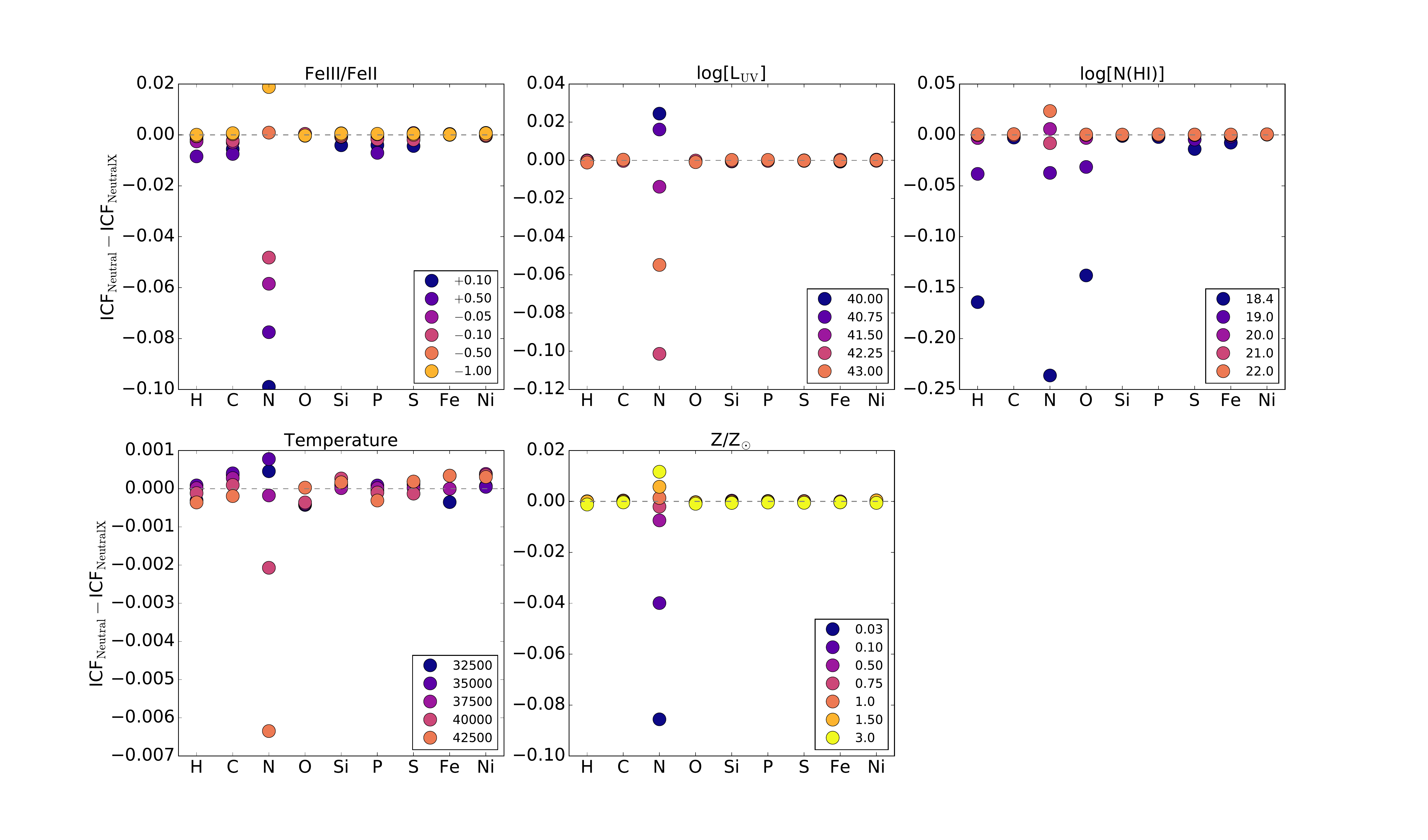}}
      \caption{Difference between the ICF$_{\rm neutral}$ estimated from the adopted standard model (listed in bold font in Table ~\ref{tab:icfs_grid}) and the rest of the models. We show the difference in ICF$_{\rm neutral}$ values as a function of element. The varying model parameter is shown in the title and legend of each of the panels. We display as a dashed grey line the ICF$_{\rm Neutral}-$ICF$_{\rm NeutralX}$ = 0.0}
         \label{fig:icfs_compare}
   \end{figure*}

In Figure \ref{fig:icfs_compare} we compare the ICF$_{\rm neutral}$ values from each of the models to those calculated from the adopted standard model. There is a general trend in all the panels of Figure \ref{fig:icfs_compare}; the largest differences in the ICF$_{\rm neutral}$ values are observed in nitrogen. Varying the $\log$[$N$(\ion{Fe}{3})/$N$(\ion{Fe}{2})], $\log$[L$_{\rm UV}$], T$_{\rm eff}$, and $Z$ has little or negligible effect in the ICF$_{\rm neutral}$ of the model grid compared to the standard model for most of the elements, except for nitrogen. For $N$(\ion{H}{1}) we find that the largest deviations from the  adopted model parameters are recorded when setting this parameter to the lowest observed value, $N$(\ion{H}{1}) = 18.4 cm$^{-2}$. However, we note that the only elements affected by this behavior are H, N and O, with differences as high as $\sim$0.25 dex. Based on Figure \ref{fig:icfs_compare} we can conclude that the high ICF$_{\rm neutral}$ values seen in nitrogen originate from a combination of factors: high \ion{Fe}{3}/\ion{Fe}{2} ratios, relatively high luminosities, low $N$(\ion{H}{1}), high T$_{\rm eff}$, and low metallicities.

\subsection{High \ion{N}{2} in the Neutral Gas}
The photoionization modeling done here shows that large amounts of \ion{N}{2} might arise from the neutral-gas regions in those cases where the ionizing source has luminosity values $\log$[L$_{\rm UV}$] $\gtrsim$ 40.75 erg s$^{-1}$. It is noteworthy to mention that although \ion{O}{1} and \ion{N}{1} have similar ionization potential, 13.6 and 14.5 eV, respectively, it is only \ion{N}{2} which appears in larger quantities in the \ion{H}{1} region. The difference in the behavior of these two elements can be explained by the fact that the ionization potential of \ion{N}{1} is 0.9 eV greater than that of \ion{H}{1} and \ion{O}{1} reducing the rate of charge exchange recombination by more than 3 orders of magnitude \citep{ost89}. This means that the ionization of \ion{N}{1} comes from a combination of contributions of photoionization and, to a lesser extent, charge exchange ionization with free protons \citep{bau99}. \par
Given that the Lyman continuum is mostly absorbed in the \ion{H}{1} cloud, we expect little photoionization from the ground state of \ion{N}{1}, therefore most of the nitrogen in the neutral gas is expected to be in \ion{N}{1}. However, the \ion{N}{1} $^{4}$S$^{\circ}$ term is only 2.4 eV above the ground state of \ion{N}{1}, easily photoionized by the Balmer continuum which typically penetrates deep into the \ion{H}{1} cloud. The critical gas volume density for this process to produce a significant population is log[$n$(H)] $\gtrsim$ 4 cm$^{-3}$ \citep[][values well within those from our galaxy sample shown in Table \ref{tab:tab_cloudy}]{est98}. Tests show that photoionization from $^{4}$S$^{\circ}$ significantly affects the \ion{N}{2} ionization in regions close to the PDR edge (Ferland priv. comm). \par
\ion{N}{2} can also be produced by photoexcitation by the Balmer continuum. \citet{bau99} proposed continuum fluorescent excitation as a critical contributor to the strong  intensity of the [\ion{N}{1}] doublet observed in objects such as Orion. Furthermore, \citet{fer12} generated detailed photoionization models to show that pumping by FUV stellar radiation might explained the anomalously strong [\ion{N}{1}] emission. This fluorescent mechanism allows for a significant fraction of excitations by FUV photons to populate the excited doublets. This is possible through two paths: 1) direct excitation by an intercombination line from the ground state, or 2) indirect excitation by a resonance line followed by de-excitation through an intercombination line \citep{fer12}. The latter route would create high column densities of \ion{N}{2} in the neutral gas, and have a fraction of that converted to [\ion{N}{1}]. This strong dependance of the fluorescent mechanism on the large flux of very hard photons or energetic particles is what we observed in the high nitrogen ICF$_{\rm neutral}$ calculated from the photoionization models with $\log$[L$_{\rm UV}$] $\gtrsim$  40.75 erg s$^{-1}$. To confirm this hypothesis we re-calculated the ICF values for I Zw 18 turning all of the fluorescent mechanisms off, which include the continuum fluorescent excitation mechanism, and found that for most of the elements the difference in the ICF values is of the order of $>$ 10$^{-5}$ dex, with the exception of nitrogen, where the ICF$_{\rm neutral}$ value goes from 0.665 dex to 0.000 dex, similar to the previously observed ICF$_{\rm neutral}$ of \ion{O}{1}.\par

\subsection{Application to systems throughout the Universe}
Considering the broad range of parameters covered in our uniform grid, it should be noted that such corrections are not limited to local star-forming galaxies and are also applicable to systems at higher redshifts. For example, the extremely compact SFGs found at redshifts of $z\sim$ 0.1-0.4 commonly known as green peas (GPs) have metallicites of $Z\sim0.20\; Z_{\odot}$ \citep{amo10}. Through a study of 17 GPs \citet{mck19} derived \ion{H}{1} column densities from the Ly$\alpha$ absorption ranging between $\log[N\rm($\ion{H}{1}$)/ \rm cm^{-2}]=$ 19-21, parameters falling within the range explored in this paper. Similarly, studies of Lyman Break galaxies (LGBs) typically observed at redshifts 2 $<z<$ 5 with metallicities ranging between $Z=$ 0.1 - 1 $Z_{\odot}$ \citep{gia02}, and \ion{H}{1} column densities of $\log[N\rm($\ion{H}{1}$)/ \rm cm^{-2}]\sim$ 21 \citep{pet00} should benefit from the ionization corrections presented here. Other typical systems observed at the peak of cosmic star formation ($z\sim$ 2-3) such as damped Ly$\alpha$ (DLA) are found to have even lower metallicities of $Z=0.01\:Z_{\odot}$ and \ion{H}{1} column densities of $\log[N\rm($\ion{H}{1}$)/\rm cm^{-2}]=$ 20-21 \citep{coo11}. All of the physical properties of the different systems listed above fall within the parameter space covered in our model grid, facilitating the calculation of ionization corrections for all those targets irrespective of redshifts.  \par

Studies of the neutral gas at high redshifts, $z\gtrsim$ 0.1, using the higher sensitivity of the next generation telescopes, i.e. James Webb Space Telescope, Extremely Large Telescope, can benefit from the ICF estimated here. Presently, ICFs are primarily attempted for ISM studies of DLA systems. For comparison \citet{coo11} estimated ICF$_{\rm ionized}$ of the order of $\lesssim$ 0.1 dex for DLAs with \ion{H}{1} column densities between $20.0 < \log[N\rm ($\ion{H}{1}$)/\rm cm^{-2}] < 21.0$, such corrections are similar to those estimated here for targets with comparable \ion{H}{1} column densities reinforcing the applicability of our model grid ICFs. 

\section{Concluding Remarks}\label{sec:con}
In studies focused on the chemical abundances of the neutral gas it is critical to take into account ionization effects due to contaminating ionized gas or unaccounted ionization stages in the neutral gas along the line of sight. In the work described here we calculate appropriate ionization corrections to the measured neutral-gas abundances in a variety of galactic environments. We infer ICFs for a set of SFGs observed with COS on board Hubble. We use the spectroscopic observations to accurately measure the column densities of \ion{Fe}{2} and \ion{Fe}{3}, using the \ion{Fe}{3}/\ion{Fe}{2} ratio as an accurate indicator of the gas volume density of the observed targets. With this additional property we then obtain ad-hoc photoionization models which allow us to quantify the corrections required for each of the SFGs analyzed here, solely based on their specific physical properties. The models show that the highest ICFs were required for one of the most extreme targets, M83-2, with the highest metallicity in our sample. \par
We also identify a trend in the ICF$_{\rm neutral}$ where the nitrogen corrections appear to be relatively higher than those for the rest of the elements. Our analysis indicated that there is a threshold in luminosity above which the ICF$_{\rm neutral}$ for nitrogen begin to increase to larger values than previously documented. We propose that these high ICF$_{\rm neutral}$ for nitrogen are due to fluorescent mechanisms producing high column densities of  \ion{N}{2} through photoexcitation by the Balmer continuum. This excess of \ion{N}{2} in the neutral gas of some of the star-forming galaxies in our sample can be confirmed through the analysis of new COS observations covering the \ion{N}{2} $\lambda$1084 \r{A} line. We note that future neutral-gas studies of luminous ionizing sources ($\log$[L$_{\rm UV}$] $\gtrsim$  40.75 erg s$^{-1}$) should carefully account for the expected \ion{N}{2} in the neutral gas cloud to accurately measure the nitrogen abundances in this neutral component. \par
Additionally, as part of our analysis we produce a grid of models that cover a wide range of physical properties of the neutral-gas environment. Using these models we make available ionization correction factors that can be applied to abundance studies of the neutral-gas under a broad range of conditions that are widely observed by the extragalactic community.

\acknowledgments

The authors want to thank the anonymous referee for comments that greatly improved the quality of this publication, as well as Max Pettini for insightful feedback. These data are associated with the HST GO programs 15193 (PI: A. Aloisi) and is available via the MAST archive \dataset[10.17909/t9-228a-7p88]{https://doi.org/10.17909/t9-228a-7p88}. Support for this program was provided by NASA through grants from the Space Telescope Science Institute. Some of the data presented in this paper were obtained from the Mikulski Archive at the Space Telescope Science Institute (MAST). The work presented here made use of the NASA/IPAC Extragalactic Database (NED), which is operated by the Jet Propulsion Laboratory, California Institute of Technology, under contract with the National Aeronautics and Space Administration.

%

\vspace{5mm}
\facilities{HST(COS)}
\software{CLOUDY \citep{fer17}, VoigtFit v.0.10.3.3 \citep{kro18}, CALCOS pipeline (v.3.3.4)}

%



\appendix

\begin{table*}
\caption{ICF$_{\rm ionized}$ for uniform grid}
\label{tab:icfs_val_grid}
\centering 
\begin{tabular}{cccccccccc}
\hline \hline
\multicolumn{10}{c}{ICF$_{\rm ionized}$}\\
 \hline
Model & \ion{H}{1} & \ion{C}{2} & \ion{N}{1} & \ion{O}{1} & \ion{Si}{2} & \ion{P}{2} & \ion{S}{2}& \ion{Fe}{2}& \ion{Ni}{2}\\
\hline\
        Standard & 0.003 & 0.095 & 0.001 & 0.003 & 0.070 & 0.098 & 0.048 & 0.022 & 0.031  \\
 \hline
	\ion{Fe}{3}/\ion{Fe}{2} = $+$0.10 & 0.005 & 0.236 & 0.001 & 0.005 & 0.125 & 0.174 & 0.069 & 0.029 & 0.034 \\ 
	\ion{Fe}{3}/\ion{Fe}{2} = $+$0.50 & 0.005 & 0.330 & 0.002 & 0.005 & 0.149 & 0.196 & 0.073 & 0.032 & 0.034 \\
	\ion{Fe}{3}/\ion{Fe}{2} = $-$0.05 & 0.004 & 0.192 & 0.001 & 0.005 & 0.111 & 0.157 & 0.065 & 0.028 & 0.034 \\
	\ion{Fe}{3}/\ion{Fe}{2} = $-$0.10 & 0.004 & 0.178 & 0.001 & 0.005 & 0.106 & 0.151 & 0.063 & 0.027 & 0.034 \\
	\ion{Fe}{3}/\ion{Fe}{2} = $-$0.50 & 0.003 & 0.092 & 0.001 & 0.003 & 0.068 & 0.096 & 0.047 & 0.022 & 0.031 \\
	\ion{Fe}{3}/\ion{Fe}{2} = $-$1.00 & 0.002 & 0.039 & 0.001 & 0.002 & 0.034 & 0.043 & 0.028 & 0.015 & 0.025 \\
	
  \hline
	$\log$[L$_{\rm UV}$] = 40.00 & 0.003 & 0. 97 & 0.002 & 0.004 & 0.080 & 0.098 & 0.047 & 0.026 & 0.030 \\
	$\log$[L$_{\rm UV}$] = 40.75 & 0.003 & 0.096 & 0.001 & 0.003 & 0.074 & 0.098 & 0.048 & 0.024 & 0.031 \\
	$\log$[L$_{\rm UV}$] = 41.50 & 0.003 & 0.095 & 0.001 & 0.003 & 0.068 & 0.099 & 0.048 & 0.021 & 0.032 \\
	$\log$[L$_{\rm UV}$] = 42.25 & 0.003 & 0.112 & 0.001 & 0.004 & 0.070 & 0.113 & 0.053 & 0.020 & 0.033 \\
	$\log$[L$_{\rm UV}$] = 43.00 & 0.002 & 0.070 & 0.001 & 0.003 & 0.043 & 0.075 & 0.041 & 0.014 & 0.031 \\
   \hline
	$\log[N\rm($\ion{H}{1}$)]=$ 18.4 & 0.339 & 0.794 & 0.318 & 0.363 & 0.789 & 0.805 & 0.774 & 0.703 & 0.805 \\
	$\log[N\rm($\ion{H}{1}$)]=$ 19.0 & 0.076 & 0.392 & 0.055 & 0.087 & 0.386 & 0.405 & 0.367 & 0.301 & 0.399 \\
	$\log[N\rm($\ion{H}{1}$)]=$ 20.0 & 0.010 & 0.144 & 0.005 & 0.012 & 0.132 & 0.157 & 0.112 & 0.068 & 0.111 \\
	$\log[N\rm($\ion{H}{1}$)]=$ 21.0 & 0.002 & 0.084 & 0.001 & 0.002 & 0.052 & 0.076 & 0.032 & 0.014 & 0.018 \\
	$\log[N\rm($\ion{H}{1}$)]=$ 22.0 & 0.000 & 0.024 & 0.000 & 0.000 & 0.009 & 0.012 & 0.004 & 0.002 & 0.002 \\
   \hline 
	T$_{\rm eff}$ = 32,500 & 0.003 & 0.116 & 0.001 & 0.003 & 0.070 & 0.100 & 0.060 & 0.025 & 0.032 \\
	T$_{\rm eff}$ = 35,000 & 0.003 & 0.106 & 0.001 & 0.003 & 0.070 & 0.098 & 0.054 & 0.024 & 0.032 \\
	T$_{\rm eff}$ = 37,500 & 0.003 & 0.097 & 0.001 & 0.003 & 0.070 & 0.098 & 0.049 & 0.022 & 0.031 \\
	T$_{\rm eff}$ = 40,000 & 0.003 & 0.088 & 0.001 & 0.003 & 0.072 & 0.100 & 0.045 & 0.021 & 0.031 \\
	T$_{\rm eff}$ = 42,500 & 0.003 & 0.080 & 0.001 & 0.003 & 0.076 & 0.105 & 0.042 & 0.020 & 0.030 \\
   \hline 
	$Z$ = 0.03 & 0.003 & 0.083 & 0.001 & 0.003 & 0.033 & 0.097 & 0.047 & 0.011 & 0.037 \\
 	$Z$ = 0.10 & 0.003 & 0.085 & 0.001 & 0.003 & 0.036 & 0.097 & 0.047 & 0.012 & 0.036 \\
	$Z$ = 0.50 & 0.003 & 0.091 & 0.001 & 0.003 & 0.057 & 0.097 & 0.047 & 0.018 & 0.033 \\
	$Z$ = 0.75 & 0.003 & 0.094 & 0.001 & 0.003 & 0.067 & 0.099 & 0.048 & 0.021 & 0.031 \\
	$Z$ = 1.00 & 0.003 & 0.096 & 0.001 & 0.003 & 0.073 & 0.099 & 0.048 & 0.023 & 0.031 \\
	$Z$ = 1.50 & 0.003 & 0.100 & 0.001 & 0.004 & 0.084 & 0.100 & 0.049 & 0.027 & 0.029 \\
	$Z$ = 3.00 & 0.003 & 0.104 & 0.001 & 0.003 & 0.094 & 0.097 & 0.049 & 0.033 & 0.024 \\

 \hline
\end{tabular}
\end{table*}

\begin{table*}
\caption{ICF$_{\rm neutral}$ for uniform grid}
\label{tab:icfs_val_grid_neutral}
\centering 
\begin{tabular}{cccccccccc}
\hline \hline
\multicolumn{10}{c}{ICF$_{\rm neutral}$}\\
 \hline
Model & \ion{H}{2} & \ion{C}{3} & \ion{N}{2} & \ion{O}{2} & \ion{Si}{3} & \ion{P}{3} & \ion{S}{3}& \ion{Fe}{3}& \ion{Ni}{3}\\
\hline\
        Standard & 0.001 & 0.001 & 0.027 & 0.000 & 0.001 & 0.000 & 0.001 & 0.000 & 0.001  \\
 \hline
	\ion{Fe}{3}/\ion{Fe}{2} = $+$0.10 & 0.003 & 0.006 & 0.126 & 0.000 & 0.005 & 0.005 & 0.005 & 0.000 & 0.000 \\ 
	\ion{Fe}{3}/\ion{Fe}{2} = $+$0.50 & 0.009 & 0.008 & 0.104 & 0.000 & 0.000 & 0.008 & 0.000 & 0.000 & 0.000\\
	\ion{Fe}{3}/\ion{Fe}{2} = $-$0.05 & 0.004 & 0.004 & 0.085 & 0.000 & 0.001 & 0.001 & 0.002 & 0.000 & 0.001 \\
	\ion{Fe}{3}/\ion{Fe}{2} = $-$0.10 & 0.002 & 0.003 & 0.075 & 0.000 & 0.001 & 0.002 & 0.003 & 0.000 & 0.001 \\
	\ion{Fe}{3}/\ion{Fe}{2} = $-$0.50 & 0.001 & 0.001 & 0.026 & 0.001 & 0.001 & 0.000 & 0.001 & 0.000 & 0.001 \\
	\ion{Fe}{3}/\ion{Fe}{2} = $-$1.00 & 0.001 & 0.000 & 0.008 & 0.001 & 0.000 & 0.000 & 0.000 & 0.000 & 0.000 \\
	
  \hline
	$\log$[L$_{\rm UV}$] = 40.00 & 0.002 & 0.001 & 0.002 & 0.001 & 0.001 & 0.001 & 0.001 & 0.001 & 0.001 \\
	$\log$[L$_{\rm UV}$] = 40.75 & 0.001 & 0.001 & 0.011 & 0.001 & 0.001 & 0.001 & 0.001 & 0.001 & 0.001 \\
	$\log$[L$_{\rm UV}$] = 41.50 & 0.001 & 0.001 & 0.041 & 0.001 & 0.001 & 0.001 & 0.001 & 0.000 & 0.000 \\
	$\log$[L$_{\rm UV}$] = 42.25 & 0.001 & 0.001 & 0.128 & 0.001 & 0.001 & 0.001 & 0.001 & 0.000 & 0.001 \\
	$\log$[L$_{\rm UV}$] = 43.00 & 0.002 & 0.000 & 0.082 & 0.001 & 0.000 & 0.000 & 0.001 & 0.001 & 0.001 \\
   \hline
	$\log[N\rm($\ion{H}{1}$)]=$ 18.4 & 0.165 & 0.003 & 0.263 & 0.138 & 0.002 & 0.003 & 0.015 & 0.008 & 0.001 \\
	$\log[N\rm($\ion{H}{1}$)]=$ 19.0 & 0.039 & 0.001 & 0.064 & 0.032 & 0.001 & 0.001 & 0.005 & 0.002 & 0.000 \\
	$\log[N\rm($\ion{H}{1}$)]=$ 20.0 & 0.004 & 0.000 & 0.021 & 0.003 & 0.001 & 0.000 & 0.001 & 0.001 & 0.000 \\
	$\log[N\rm($\ion{H}{1}$)]=$ 21.0 & 0.001 & 0.001 & 0.035 & 0.000 & 0.001 & 0.001 & 0.000 & 0.000 & 0.000 \\
	$\log[N\rm($\ion{H}{1}$)]=$ 22.0 & 0.001 & 0.000 & 0.004 & 0.000 & 0.000 & 0.000 & 0.000 & 0.000 & 0.000 \\
   \hline 
	T$_{\rm eff}$ = 32,500 & 0.001 & 0.000 & 0.026 & 0.001 & 0.001 & 0.001 & 0.001 & 0.001 & 0.001\\
	T$_{\rm eff}$ = 35,000 & 0.001 & 0.000 & 0.026 & 0.001 & 0.001 & 0.000 & 0.001 & 0.000 & 0.001 \\
	T$_{\rm eff}$ = 37,500 & 0.001 & 0.001 & 0.027 & 0.001 & 0.001 & 0.001 & 0.001 & 0.000 & 0.000 \\
	T$_{\rm eff}$ = 40,000 & 0.001 & 0.001 & 0.029 & 0.001 & 0.000 & 0.001 & 0.001 & 0.000 & 0.000 \\
	T$_{\rm eff}$ = 42,500 & 0.001 & 0.001 & 0.033 & 0.000 & 0.001 & 0.001 & 0.001 & 0.000 & 0.001 \\
   \hline 
	$Z$ = 0.03 & 0.001 & 0.000 & 0.113 & 0.001 & 0.001 & 0.000 & 0.001 & 0.000 & 0.001 \\
 	$Z$ = 0.10 & 0.001 & 0.000 & 0.067 & 0.001 & 0.000 & 0.000 & 0.001 & 0.000 & 0.001 \\
	$Z$ = 0.50 & 0.001 & 0.001 & 0.034 & 0.001 & 0.001 & 0.000 & 0.001 & 0.001 & 0.000 \\
	$Z$ = 0.75 & 0.001 & 0.001 & 0.029 & 0.001 & 0.001 & 0.001 & 0.001 & 0.000 & 0.000 \\
	$Z$ = 1.00 & 0.002 & 0.001 & 0.026 & 0.001 & 0.001 & 0.001 & 0.001 & 0.001 & 0.000 \\
	$Z$ = 1.50 & 0.001 & 0.001 & 0.021 & 0.001 & 0.001 & 0.001 & 0.001 & 0.001 & 0.000 \\
	$Z$ = 3.00 & 0.002 & 0.001 & 0.015 & 0.001 & 0.001 & 0.001 & 0.001 & 0.001 & 0.001 \\

 \hline
\end{tabular}
\end{table*}



\bibliographystyle{aasjournal}
\bibliography{SFGs} 


\end{document}